\documentclass[12pt,a4paper,oneside]{article}
\pdfoutput=1 
\usepackage{graphicx}
\usepackage{sectsty,amssymb,amsmath,color}
\usepackage[linkcolor={blue},citecolor={red},colorlinks=true]{hyperref}
\usepackage{enumerate}
\usepackage[margin=2cm]{geometry}
\usepackage{multirow}
\usepackage{authblk}
\usepackage{slashed}
\usepackage[sort]{cite}
\usepackage{soul}
\usepackage{color}
\def\beq{\begin{equation}}
\def\eeq{\end{equation}}
\def\bea{\begin{eqnarray}}
\def\eea{\end{eqnarray}}

\def\bit{\begin{itemize}}
\def\eit{\end{itemize}}

\def \l{\left}
\def \r{\right}

\def\ra{\rightarrow}

\def\baa{\begin{array}}
\def\eaa{\end{array}}

\def\sl#1{\mathord{\not\mathrel{{\mathrel{#1}}}}}

\def\simgt{\mathrel{\lower2.5pt\vbox{\lineskip=0pt\baselineskip=0pt
           \hbox{$>$}\hbox{$\sim$}}}}
\def\simlt{\mathrel{\lower2.5pt\vbox{\lineskip=0pt\baselineskip=0pt
           \hbox{$<$}\hbox{$\sim$}}}}

\definecolor{green1}{rgb}{0.0, 0.5, .0}

\newcommand{\hhreflink}[1]{\href{http://cds.cern.ch/record/1611186}{\color{cyan}{#1}}}
\newcommand{\hhrefA}[1]{\href{http://arxiv.org/abs/#1}{\color{cyan}{arXiv:#1}}}
\newcommand{\hhrefhp}[1]{\href{http://arxiv.org/abs/hep-ph/#1}{\color{cyan}{arXiv:hep-ph/#1}}}

\begin{document}

\begin{center}
{\Large \bf  
 {Resolving gluon fusion loops}
}
\end{center}
\begin{center}
{\Large \bf  
 {at current and future hadron colliders}
}
\end{center}
\vskip0.5cm

\renewcommand{\thefootnote}{\fnsymbol{footnote}}
\begin{center}
{\large Aleksandr Azatov,$^{a}$ Christophe Grojean,$^{b\,}$\footnote{\hspace{0.1cm}On  leave  of absence from  ICREA,  E-08010  Barcelona,  Spain and IFAE, Barcelona Institute of Science and Technology (BIST) Campus UAB, E-08193 Bellaterra, Spain.} Ayan Paul$^{c}$ and Ennio Salvioni$^{d}\,$
\footnote{Email: aleksandr.azatov@sissa.it, christophe.grojean@desy.de, apaul2@alumni.nd.edu and \\ennio.salvioni@tum.de.}
}
\end{center}
\begin{center}
%
\centerline{$^{a}${\small \it Abdus Salam International Centre for Theoretical Physics, I-34151 Trieste, Italy}}
\vskip 4pt
%
\centerline{$^{b}${\small \it DESY, D-22607 Hamburg, Germany
}}
\vskip 4pt
%
\centerline{$^{c}${\small \it INFN, Sezione di Roma, I-00185 Rome, Italy}}
\vskip 4pt
%
\centerline{$^{d}${\small \it Physics Department, University of California, Davis, CA 95616, USA}}
\end{center}

\renewcommand{\thefootnote}{\arabic{footnote}}
\setcounter{footnote}{0}

\vglue 1.0truecm

\begin{abstract}
Inclusive Higgs measurements at the LHC have limited resolution on the gluon fusion loops, being unable to distinguish the long-distance contributions mediated by the top quark from possible short-distance new physics effects. Using an Effective Field Theory (EFT) approach we compare several proposed methods to lift this degeneracy, including $t\bar{t}h$ and boosted, off-shell and double Higgs production, and perform detailed projections to the High-Luminosity LHC and a future hadron collider. In addition, we revisit off-shell Higgs production. Firstly, we point out its sensitivity to modifications of the top-$Z$ couplings, and by means of a general analysis we show that the reach is comparable to that of tree-level processes such as $t\bar{t}Z$ production. Implications for composite Higgs models are also discussed. Secondly, we assess the regime of validity of the EFT, performing an explicit comparison for a simple extension of the Standard Model containing one vector-like quark.
\end{abstract}

\newpage
\section{Introduction}
One of the main goals of the Large Hadron Collider (LHC) is to unveil the origin of the electroweak symmetry breaking (EWSB): is it driven by a solitary and elementary Higgs field as in the Standard Model (SM), or is there additional dynamics not too far above the weak scale? New physics around the TeV frontier can reveal itself in a direct way, through the discovery of new particle resonances, or indirectly, via modifications of the interactions of the SM fields.

Since the discovery of the Higgs boson in 2012, the numerous LHC measurements aimed at testing its properties have revealed an increasingly precise profile consistent with the SM predictions~\cite{Aad:2015zhl,PDG2016}. Yet, no information can be extracted on the values of the Higgs couplings without assumptions, for instance on the Higgs boson total width. Furthermore, the current measurements, being dominated by inclusive observables, suffer from `blind' directions in the exploration of the parameter space of the Higgs couplings. 

In particular, as emphasized in Refs.~\cite{Azatov:2013xha,Grojean:2013nya}, the current constraints allow for $O(1)$ deviations of the $h\bar{t}t$ coupling if correlated contact interactions between the Higgs boson and gluons and photons are simultaneously present. Far from a mere academic question, this degeneracy is especially relevant in models where the Higgs is a composite pseudo Nambu-Goldstone boson (pNGB)~\cite{Bellazzini:2014yua,Panico:2015jxa}, where the inclusive Higgs rates are typically insensitive to the spectrum of the fermionic resonances~\cite{Falkowski:2007hz,Low:2010mr,Azatov:2011qy,Delaunay:2013iia,Montull:2013mla}. An analogous situation can be realized in natural supersymmetry, where the top and stop loops can conspire to leave the inclusive Higgs production SM-like~\cite{Grojean:2013nya}. In these scenarios indirect signs of the top partners, which play a crucial role in addressing Higgs naturalness, can therefore only be seen by accessing individually the $h\bar{t}t$ and $hgg$ couplings in exclusive measurements. The most obvious candidate is Higgs production in association with a top quark pair, see for example Refs.~\cite{project,Plehn:2015cta} for recent studies. However, in the last few years several other proposals have been put forward, including boosted Higgs production~\cite{Harlander:2013oja,Banfi:2013yoa,Azatov:2013xha,Grojean:2013nya,Schlaffer:2014osa,Buschmann:2014twa,Langenegger:2015lra,Grazzini:2015gdl} (see also Refs.~\cite{Langenegger:2006wu,Arnesen:2008fb,Bagnaschi:2011tu} for previous studies where the Higgs transverse momentum distribution was exploited as a handle on new physics), off-shell Higgs production~\cite{Cacciapaglia:2014rla,Azatov:2014jga,Buschmann:2014sia}, and double Higgs production in gluon fusion~\cite{Azatov:2015oxa,Goertz:2014qta}. In Section~\ref{sec:projections} of this paper we combine existing results for all the above processes, to estimate the future resolution on the Higgs gluon fusion loops at the High-Luminosity LHC (HL-LHC), defined as a $14$\,TeV $pp$ collider with $3$\,ab$^{-1}$ of integrated luminosity, and at the hadron-hadron Future Circular Collider (FCC-hh, abbreviated to FCC except where confusion is possible with the electron-positron version, FCC-ee), defined as a $100$\,TeV $pp$ collider with benchmark integrated luminosity of $20$\,ab$^{-1}$. 

Our projections are presented in the context of an effective field theory (EFT) with only two dimension-$6$ operators, one parameterizing the $h\bar{t}t$ coupling and the other the $hgg$ (and $h\gamma\gamma$) contact interaction. This relies on the assumption that all the other dimension-$6$ operators will be bounded to much higher accuracy by inclusive measurements, and would thus have a negligible effect on our results. In Section~\ref{sec:ttZeffects}, however, we reconsider this assumption for off-shell Higgs production. We first observe that modifications of the top-$Z$ couplings, which affect the $gg\to ZZ$ process through top box diagrams, will be constrained at the HL-LHC with relatively low accuracy~\cite{Rontsch:2014cca,Dror:2015nkp,Bylund:2016phk} and can therefore affect the off-shell measurement in a significant way. In fact, by performing a detailed analysis we show that $gg\to ZZ$ can test the top-$Z$ couplings with a sensitivity comparable to tree-level measurements, such as $t\bar{t}Z$ production (a similar conclusion was recently obtained for $gg\to hZ$ in Refs.~\cite{Bylund:2016phk,Englert:2016hvy}). This is especially interesting in composite pNGB models, where corrections to the top-$Z$ and top-Higgs couplings can have comparable size. 

The EFT interpretation of measurements that probe a broad energy range, such as the boosted, off-shell and double Higgs productions, requires special care to ensure consistency, as discussed for example in Ref.~\cite{Contino:2016jqw}. In Section~\ref{sec:EFTvalidity} we scrutinize this aspect for off-shell Higgs production. To test the validity of the EFT we employ a toy model with a new vector-like quark, which captures the important features of more complete ultraviolet (UV) constructions, while at the same time allowing us to compare the full and EFT constraints without unnecessary complications. 

The paper is then concluded in Section~\ref{sec:summary} by a summary of our main results, as well as some comments on the outlook. A pair of appendices provide the technical details of the off-shell and boosted Higgs analyses.

\section{HL-LHC and FCC prospects} 
\label{sec:projections}
%
We begin by reviewing the modifications of Higgs production through gluon fusion in the presence of new physics interactions. In the rest of the paper we assume that there is a mass gap between the SM states and new resonances, so that the electroweak symmetry is linearly realized and all beyond-the-SM (BSM) effects can be consistently parameterized in term of higher-dimensional operators. Operators that can modify the Higgs production through gluon fusion first appear at dimension $6$. In this section we consider only the following subset
\bea \label{eq:L6}
 {\cal L}_{6}& = &c_y \frac{y_t |H|^2}{v^2} \bar{Q}_L \widetilde{H} t_R+\mathrm{h.c.}+\frac{c_gg_s^2}{48\pi^2 v^2}|H|^2G_{\mu\nu}G^{\mu\nu}, 
\eea
which after EWSB modify the interactions between the Higgs boson and the top quark and gluons,
\bea
\label{eq:nonl}
{\cal L}_{\mathrm{nl}}&= &-\,c_t \,\frac{m_t}{v}\, \bar{t} t h + \frac{c_g g_s^2}{48\pi^2}\frac{h}{v} G_{\mu\nu}G^{\mu\nu}\,,\qquad  c_t=1-c_y\,.
\eea
While several other operators affect Higgs physics (see for instance Refs.~\cite{Giudice:2007fh, Contino:2013kra}), we choose to focus only on those in Eq.~\eqref{eq:L6} because the determination of their coefficients is plagued by a well-known degeneracy in the fit to inclusive Higgs data. In fact, the Higgs Low Energy Theorem (LET)~\cite{Shifman:1979eb,Ellis:1975ap} tells us that to good approximation, the total Higgs production is sensitive only to the linear combination $c_g + c_t$, and is thus blind along the line $\left|c_t + c_g\right| = {\rm constant}$. In addition, while the $h\to \gamma\gamma$ decay width depends on $c_t$ via top loops, if the contact operator in Eq.~\eqref{eq:L6} is mediated by states with top-like SM unbroken quantum numbers (electric charge equal to $2/3$ and fundamentals of color), then in addition to Eq.~\eqref{eq:nonl} the following effective coupling is generated
\bea
c_g\,\frac{e^2}{18\pi^2}\frac{h}{v}F_{\mu\nu}F^{\mu\nu}\,.
\label{eq:photon}
\eea
In this case the $h\to \gamma\gamma$ amplitude again depends on the linear combination $c_t + c_g$. The choice of top-like quantum numbers for the new fields is strongly motivated by models addressing the hierarchy problem, namely composite Higgs and natural supersymmetry. Under this compelling assumption, the inclusive Higgs measurements cannot resolve the degeneracy between $c_t$ and $c_g$.   

\begin{figure}
\begin{center}
\includegraphics[scale=0.585]{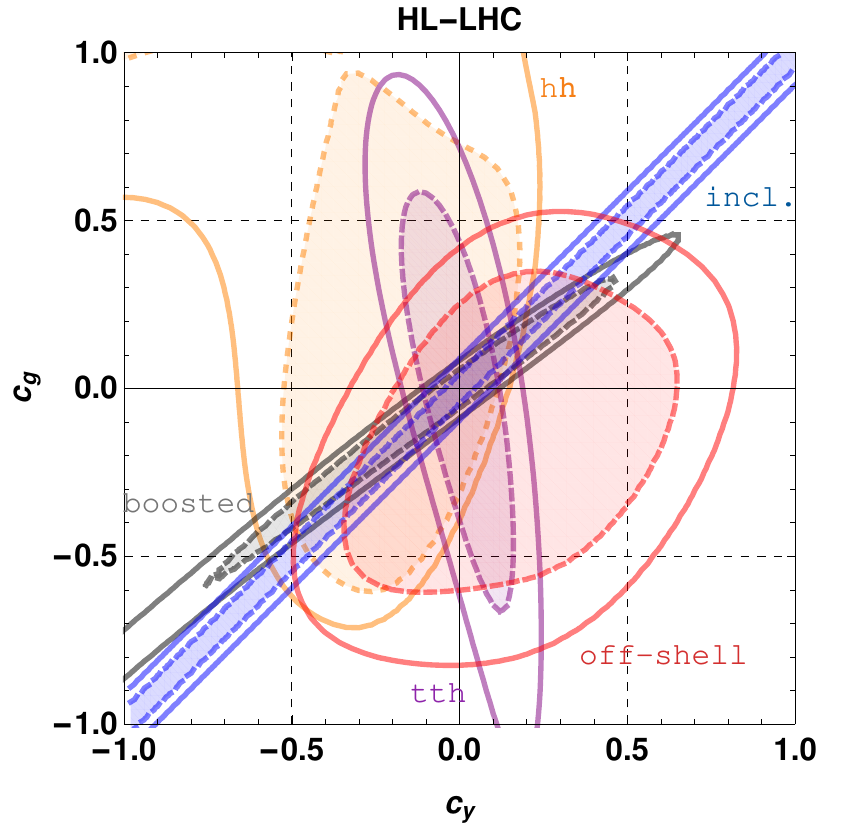}
\includegraphics[scale=0.55]{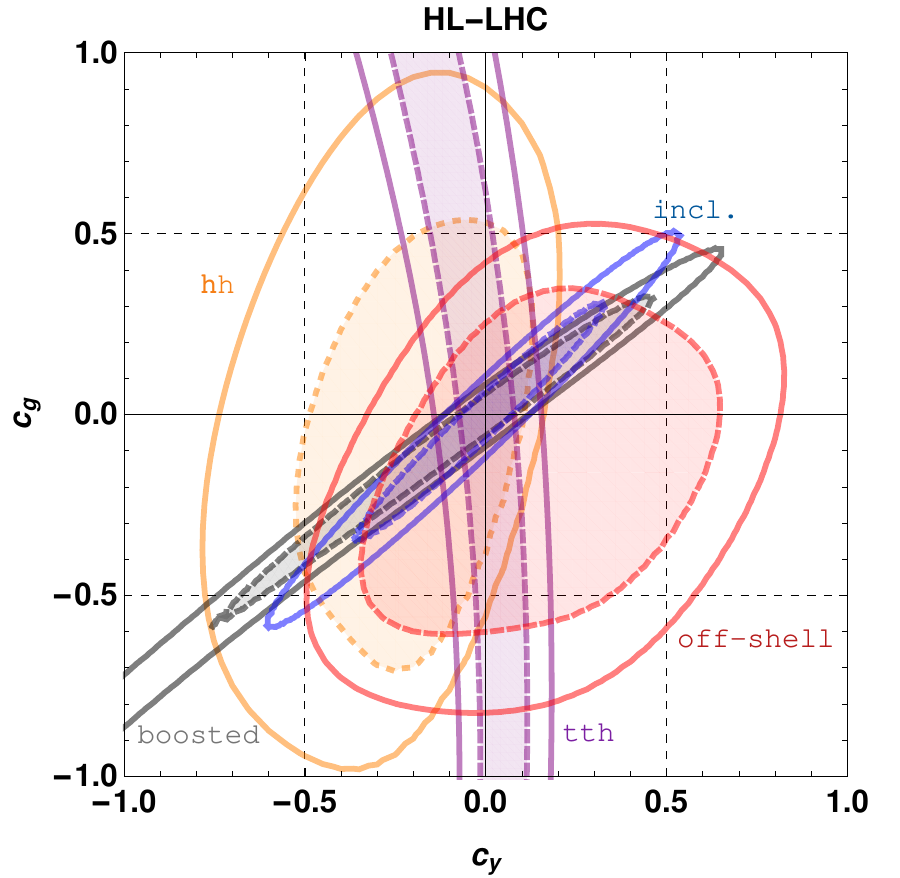}
\end{center} 
\caption{{\it Left panel:} $95\%$ (solid) and $68\%$ (dashed) exclusion contours in the $(c_y,\,c_g)$ plane obtained from HL-LHC projections: inclusive Higgs measurements (blue), $t\bar{t}h$ (purple), off-shell (red), boosted (gray), and double Higgs production (orange). The $h\gamma\gamma$ effective interaction in Eq.~\eqref{eq:photon} is included. {\it Right panel:} Same as in the left panel, but without the $h\gamma\gamma$ effective interaction in Eq.~\eqref{eq:photon}.\protect\footnotemark}
\label{fig:comb}
\end{figure}

Nevertheless, a few exclusive measurements have the potential to break this degeneracy, by individually accessing $c_t$ and $c_g$. The aim of this section is to give the projected sensitivity at the HL-LHC and FCC for each of these channels. We begin by summarizing the measurements and how our projections were derived:
\bit
\item 
{\bf Higgs and top quark pair associated production:} This is the only channel among those we consider that probes $c_t$ at tree level. The signal rate is proportional to $|c_t|^2$, with some minor dependence on $c_g$ mainly coming from the modification of the total Higgs width, and to a lesser extent from the additional diagrams contributing to $t\bar{t}h$ production~\cite{Degrande:2012gr}. We estimate the reach at the HL-LHC using the ATLAS study in Ref.~\cite{project}, which we recast to obtain projected exclusion contours in the $(c_y,c_g)$ plane. The sensitivity mainly comes from the decay channels $h\to ZZ, \gamma\gamma$ and $\mu\mu$, for all of which the relative uncertainty on the signal strength modifier is expected to be $\sim 20\%$ after including systematics (see Table~17 of Ref.~\cite{project}, third column). For the FCC we use instead the results of Ref.~\cite{Plehn:2015cta}.
\item {\bf Boosted Higgs production:} In this process the Higgs boson is produced in association with a QCD jet. If the jet is hard enough, $p_T\gtrsim m_t$, the parameterization of the top loops as point-like interactions between the Higgs and the gluons is invalidated. In this kinematic region the cross section becomes sensitive to $c_t$ and $c_g$ separately, providing a handle to differentiate between the two couplings~\cite{Azatov:2013xha,Grojean:2013nya}. For the HL-LHC projection we adapt the results presented in Ref.~\cite{Schlaffer:2014osa}, focusing on the $h\to \tau\tau$ channel, which was found to be the most promising \cite{Schlaffer:2014osa}. For the FCC, since no $100$\,TeV analysis is currently available, we rescale the results of Ref.~\cite{Schlaffer:2014osa} by using parton luminosity ratios. Details on the procedure, as well as the results, are given in Appendix~\ref{sec:hj}.
\item {\bf Off-shell Higgs production:} In the process $gg\to ZZ\to 4\ell$, Higgs production can be probed far off-shell, at partonic center of mass energies $\sqrt{\hat{s}}\gtrsim m_t$. Similar to the boosted Higgs production, in this kinematic regime the top quark loops cannot be parameterized by point-like interactions, so the $4\ell$ invariant mass distribution can resolve $c_t$ from $c_g$ as advocated in Ref.~\cite{Azatov:2014jga},\footnote{Notice that the inclusion of angular correlations was shown to improve the sensitivity of the off-shell Higgs analysis~\cite{Buschmann:2014sia}, but to be conservative in this paper we focus only on the $4\ell$ invariant mass distribution.} see also Ref.~\cite{Cacciapaglia:2014rla}. We estimate the HL-LHC and FCC prospects by means of the results provided in Ref.~\cite{Azatov:2014jga}, considering only statistical uncertainties.
\item {\bf Double Higgs production in gluon fusion:} The interest of this channel is twofold. Firstly, it occurs at energies larger than the top quark mass, so the point-like Higgs-gluon interactions mediated by UV physics and the top loops lead to different effects, in analogy with the previous two processes. Secondly, under our assumption that the Higgs boson belongs to an $SU(2)_L$ doublet, additional contact interactions involving two Higgses are predicted by the EFT,
\bea
{\cal L}_{\rm nl}^{hh}& = & -\, \frac{m_t}{v} \bar t t \l( c_t h + c_{2t} \frac{h^2}
{v}\r)+\frac{c_g g_s^2}{48\pi^2 }\l(\frac{h}{v}+\frac{h^2}{2 v^2}\r) G_{\mu\nu}G^{\mu\nu}\,,\qquad c_{2t}=-\frac{3}{2}c_y\,,
\eea
where $c_t$ was defined in Eq.~\eqref{eq:nonl}. These higher-point interactions make double Higgs production especially sensitive to the top Yukawa sector~\cite{Dib:2005re,Grober:2010yv,Contino:2012xk,Gillioz:2012se}. Recent studies that derived constraints on the top-Higgs interactions from double Higgs production can be found in Refs.~\cite{Azatov:2015oxa,Goertz:2014qta}. In this paper we use the results of Ref.~\cite{Azatov:2015oxa}, based on the $b\bar{b}\gamma\gamma$ final state, to estimate the HL-LHC and FCC reach.\footnote{We are grateful to R.~Contino, G.~Panico and M.~Son for providing us with the exact likelihood of the analysis in Ref.~\cite{Azatov:2015oxa}.\\\\} Only statistical uncertainties are included.
\eit
The projections for the HL-LHC are shown in the left panel of Fig.~\ref{fig:comb}, where exclusion contours in the $(c_y,\,c_g)$ plane were drawn under the assumption that data agree with the SM predictions in all channels. The additional $h\gamma\gamma$ contact interaction of Eq.~\eqref{eq:photon} was assumed to be present, while all the other Higgs couplings were assumed to have their SM values.
\footnotetext{In the previous version of the paper the contours for double Higgs production were incorrect, due to a mistake in the normalization of the event yields used. This applied to both panels.}
As a result the constraints from inclusive Higgs measurements, which were derived using the ATLAS study of Ref.~\cite{project},\footnote{We performed a global fit assuming the projected uncertainties on the signal strengths reported in the third column of Table~17 of Ref.~\cite{project}, which include systematics. All the channels except for $Z\gamma$ were included in the fit, leading to the constraint $c_g = c_y \pm 0.04\,(0.09)$ at $68\,(95)\%$ CL.} are blind along the line $c_t + c_g = 1$. Our projections show that the best channel in resolving the degeneracy is Higgs production in association with a top pair. However, the sensitivity of double Higgs production for $c_y >0,\,c_g <0$ is stronger than that of the $t\bar{t}h$ channel. This originates from the very quickly growing contributions to $gg\to hh$ coming from the diagrams with $hh gg$ and $hh \bar{t}t$ contact interactions.\footnote{If the assumption of doublet Higgs is relaxed (no $hhgg$ and $hh\bar{t}t$ interactions) the constraints become weaker.} A comment is also in order on the role of the Higgs trilinear coupling: while in our analysis it was, for definiteness, set to the SM value, even $O(1)$ departures from it would have only small effects on the results~\cite{Azatov:2015oxa}. The off-shell Higgs process yields strong constraints for $c_y < 0,\,c_g > 0$ and is therefore complementary to $hh$, whereas boosted Higgs gives a somewhat weaker bound.

For the sake of illustration, in the right panel of Fig.~\ref{fig:comb} we also present results for the scenario where the $h\gamma\gamma$ interaction in Eq.~\eqref{eq:photon} is absent. In this case the $(c_t,c_g)$ degeneracy of inclusive measurements is lifted by the $h\to \gamma\gamma$ channel, but only to a limited extent. Notice that since the $t\bar{t}h$ and $hh$ projections rely in part on the $h\to \gamma\gamma$ decay, the corresponding contours are different in the two panels, while the off-shell and boosted Higgs projections are identical, because they are based on $h^\ast \to ZZ$ and $h\to \tau\tau$, respectively.

\begin{figure}
\begin{center}
\includegraphics[scale=0.55]{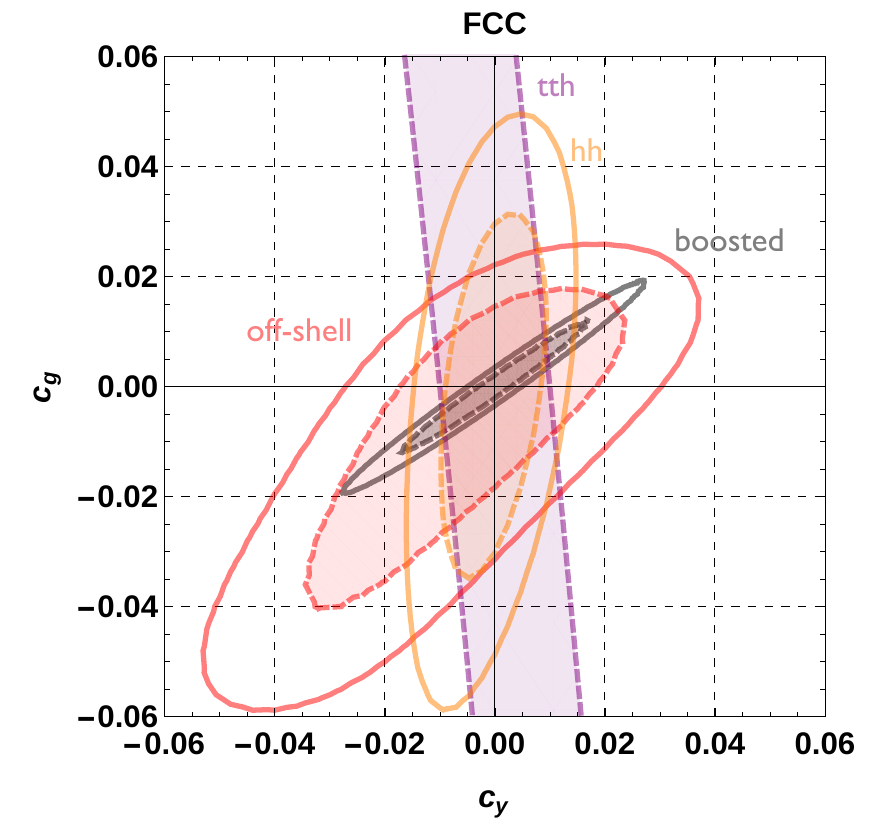}
\caption{$95\%$ (solid) and $68\%$ (dashed) exclusion contours in the $(c_y, c_g)$ plane obtained from FCC projections. Red corresponds to off-shell, gray to boosted and orange to double Higgs production, while the purple band indicates the $68\%$  region from $t \bar th$.}
\label{fig:combFCC}
\end{center}
\end{figure}

Next, we discuss the opportunities of resolving the gluon fusion loops at the FCC. We present exclusions contours in the ($c_y,\,c_g$) plane in Fig.~\ref{fig:combFCC}, where we have again assumed experimental data to agree with the SM in all channels. In addition, the $h\gamma\gamma$ effective coupling in Eq.~\eqref{eq:photon} was assumed to be present and all the other Higgs couplings were set to their SM values, leading to the insensitivity of the inclusive Higgs measurements along the $c_t + c_g = 1$ line. However, we have refrained from reporting the corresponding exclusion, because the FCC inclusive Higgs measurements will be dominated by systematics, 
and a dedicated study is currently not available. We see that the best candidates to resolve the degeneracy at the FCC are $t\bar{t}h$ and $hh$ production. Notice that, in comparison to Fig.~\ref{fig:comb}, the double Higgs contour is more closely aligned to $c_y = 0$. This happens because the SM amplitude is predominantly imaginary, whereas the piece mediated by $c_g$ is real, hence the SM-BSM interference term, which drives the constraint at the FCC, is essentially proportional to $c_y$. The alignment is less prominent at the HL-LHC, where $|$BSM$|^2$ 
terms are important because larger deviations from the SM are allowed. On the other hand, comparing with Fig.~\ref{fig:comb} we see that boosted Higgs shows a strong improvement 
at the FCC, while off-shell Higgs production is the channel that benefits the least from the increased collider energy. The reason is that the off-shell cross section, in the kinematic region $\sqrt{\hat{s}}\gtrsim 1$ TeV which becomes accessible at the FCC, contains $|\hbox{BSM}|^2$ terms that are comparable in size to the SM-BSM interference terms in the relevant region of the $(c_y, c_g)$ plane (see Eq.~\eqref{eq:yieldFCC} in Appendix~\ref{sec:simulation}). This leads to the appearance of a second distinct likelihood maximum for $c_g < 0$, which in turn implies a flattening of the full likelihood and therefore a weaker constraint. This effect is absent in the boosted Higgs measurement, where the interference term dominates the cross section (see Eq.~\eqref{eq:fcchj} in Appendix~\ref{sec:hj}) and the likelihood is sharply peaked at the SM point.

Finally, it is worthwhile to comment on other processes which, although not included in our projection, can in principle also be used to resolve the degeneracy: 
\begin{itemize}
\item $gg\to hZ$, sensitive to $c_t$ through top loops~\cite{Harlander:2013mla,Englert:2013vua}, but not to $c_g$. Recently, Refs.~\cite{Bylund:2016phk,Englert:2016hvy} found that, in analogy to our results for $gg\to ZZ$ that will be presented in Section~\ref{sec:ttZeffects}, modifications of the top-$Z$ interactions can have important effects on $gg\to hZ$. A more careful comparison of the two processes is therefore postponed to Section~\ref{sec:comparison}.
\item $pp\to t\bar{t}hh$~\cite{Englert:2014uqa}, which can also access $c_y$, in particular through the $hh\bar{t}t$ contact interaction that leads to a linear growth with energy of the amplitude~\cite{Dror:2015nkp}. Unfortunately, a dedicated study of this aspect is still missing from the literature. 
\item Four top production~\cite{Lillie:2007hd,Pomarol:2008bh}, whose sensitivity to $c_t$ was recently studied in Ref.~\cite{Cao:2016wib}. Notice that the $pp\to t\bar{t}t\bar{t}$ cross section is also affected by deviations in the top-$Z$ couplings, but to our knowledge a combined analysis has not been performed yet.
\item Associated production of a single top with a Higgs, which, however, is only sensitive to $O(1)$ deviations of $c_t$ from the SM~\cite{Farina:2012xp,Biswas:2012bd}. 
\end{itemize}
%

\section{Effects of top-$\mathbf{Z}$ couplings in off-shell Higgs}
\label{sec:ttZeffects}

The results presented in Section~\ref{sec:projections} were obtained assuming an EFT containing the two dimension-$6$ operators in Eq.~\eqref{eq:L6}. However, it is important to verify how robust this treatment is. Firstly, we should ask whether we have included all the operators that are relevant for the processes we study, and at the same time are generated in interesting BSM theories, in particular those addressing the hierarchy problem. Secondly, it is important to check (possibly after including extra operators, as per the first point) if the EFT provides a valid and accurate description of the underlying BSM physics. Clearly, a simplified model is the ideal setup for this comparison. In the remainder of the paper we address these questions in detail for off-shell Higgs production. In this section we discuss the role of additional operators, while the assessment of the validity of the EFT is the subject of Section~\ref{sec:EFTvalidity}. 

We find that operators that modify the top-$Z$ interactions, being subject to relatively mild bounds from direct measurements, can affect the box diagrams that contribute to $gg\to ZZ$ in a significant way. At the same time, these operators typically appear in composite Higgs models with a size comparable to that of $c_y$ and $c_g$. This is exemplified by a toy model with a single vector-like quark added to the SM, with which we begin our discussion. We then move on to present the main result of this section: The extension of the analysis of Ref.~\cite{Azatov:2014jga} to include corrections to the $Z\bar{t}t$ couplings, which were neglected in all previous off-shell Higgs studies. We continue with a discussion of the implications of our results for more realistic composite Higgs models, and end the section with some comments on other gluon-fusion processes that are also sensitive to the top-$Z$ interactions, $gg\to hZ$ and $gg\to WW$. 

\subsection{Toy model with a single vector-like quark} \label{subsec:toy model}
To highlight the importance of the $Z\bar{t}t$ interactions, we introduce a toy model that arguably realizes the simplest example of $(c_t,c_g)$ degeneracy. This ambiguity naturally appears~\cite{Azatov:2011qy,Montull:2013mla} in models where the SM fermion masses are generated by the partial compositeness mechanism~\cite{Kaplan:1991dc}. A very simplified version of this framework, which is nonetheless sufficient for our purpose, is obtained by extending the SM with a single vector-like quark $T$, singlet under $SU(2)_L$
\bea
\label{eq:singletop}
{\cal L} & = & -\,y \bar Q_L \tilde H t_R- Y_* \bar Q_L \tilde H T_R - M_* \bar T_L T_R + \mathrm{h.c.}\, .
\eea
Integrating out $T$ at the tree level generates the following low-energy Lagrangian (we define $s_w \equiv \sin \theta_w$, $c_w \equiv \cos\theta_w$)
\bea
{\cal L}^{\rm EFT,\,tree}& = &-\;\frac{m_t}{v}\l( 1-\frac{Y_* ^2 v^2}{2 M_*^2}\r) h\bar t t +\frac{e }{s_w c_w}\l(\frac{1}{2}-\frac{2}{3}s_w^2 - \frac{Y_*^2 v^2}{4M_*^2}\r)Z_\mu \bar t_L \gamma^\mu t_L\nonumber\\
&&+\;\frac{e }{s_w c_w}\l(-\frac{2}{3} s_w^2 \r) Z_\mu \bar t_R \gamma^\mu t_R+  O \l(\frac{1}{M_*^4}\r),
\label{eq:tpEFTtree}
\eea
whereas at $1$-loop the following additional interaction is generated
\bea \label{eq:EFTloop}
{\cal L}^{\rm EFT,\, loop}& = &\frac{g_s^2}{48\pi^2 }\frac{h}{v}G_{\mu\nu}G^{\mu\nu}\l(\frac{Y_*^2 v^2}{2 M_*^2}\r) + O\left(\frac{1}{M_\ast^4}\right).
\eea
Notice that at $1$-loop other interactions arise (for example, dipole-type couplings), however Eq.~\eqref{eq:EFTloop} is the only one that contributes to $gg \to ZZ$ without further loop suppressions. From Eqs.~(\ref{eq:tpEFTtree},\,\ref{eq:EFTloop}) we see that the model is aligned exactly along the $c_t+c_g=1$ direction,
\bea
c_t= 1-\frac{Y_* ^2 v^2}{2 M_*^2},\qquad c_g=\frac{Y_*^2 v^2}{2 M_*^2}\,.
\eea
The relation $c_t + c_g = 1$, which can also be derived by applying the Higgs LET, implies that the inclusive Higgs production rate is identical to the SM one, even though the top Yukawa coupling receives a correction proportional to the mixing with the new vector-like quark. In addition, from Eq.~\eqref{eq:tpEFTtree} we see that the interactions of the top quark with the $Z$ boson receive corrections as well. These can be parameterized by extending the dimension-$6$ Lagrangian of Eq.~\eqref{eq:L6} to
\bea
 {\cal L}_{6}^{\rm extended}& = &c_y \frac{y_t |H|^2}{v^2} \bar{Q}_L \widetilde{H} t_R+\mathrm{h.c.}+\frac{c_g g_s^2}{48\pi^2 v^2} |H|^2 G_{\mu\nu}G^{\mu\nu}
 \nonumber\\
&&+\;
\frac{i c^3_{Hq}}{v^2} H^\dagger \sigma^a D_{\mu} H\, \bar Q_L \sigma^a \gamma^\mu  Q_L + \mathrm{h.c.}
+ 
\frac{ic^1_{Hq}}{v^2} H^\dagger  D_\mu H \,\bar Q_L \gamma^\mu  Q_L + \mathrm{h.c.}\, ,
\label{eq:opH}
\eea
with effective coefficients
\bea
\label{wilsoncoeff}
c_y = c_g = \frac{Y_* ^2 v^2}{2 M_*^2}\,,\qquad  c^1_{Hq} = - c^3_{Hq} = \frac{Y_* ^2 v^2}{4 M_*^2}\,.
\eea
This simple example shows that in models that exhibit the $(c_t,c_g)$ degeneracy, BSM effects in the $h\bar{t}t$ and $hgg$ couplings can be accompanied by modifications of comparable size to the top-$Z$ interactions. This strongly motivates the extension of the off-shell Higgs analysis of Ref.~\cite{Azatov:2014jga} to include the effects of $Z\bar{t}t$ corrections in a general way, to which the next subsection is devoted.

\subsection{Off-shell Higgs analysis including top-Z couplings}
We begin by setting our notation. The top-$Z$ couplings can be parameterized as
\bea
e Z_\mu \bar t \gamma^\mu \left(c_V + c_A \gamma_5  \right) t  \; = \; Z_\mu \bar{t} \gamma^\mu \l( c_L\, g_L^{\rm SM} P_L + c_R\, g_R^{\rm SM} P_R \r) t
\eea
where $P_{L,R} = (1\mp \gamma_5)/2$ and $g_{L}^{\rm SM} = e(1/2 - 2 s^2_w/3)/(s_w c_w)$, $g_{R}^{\rm SM} = e(- 2 s^2_w/3)/(s_w c_w)$. The SM values of the parameters are 
\begin{equation}
c_V^{\rm SM} = (1-8s_w^2/3)/(4s_w c_w) \simeq 0.23\,, \quad c_A^{\rm SM} = -1/(4s_w c_w) \simeq -0.59\,,\quad c_L^{\rm SM} = c_R^{\rm SM} = 1\,,
\end{equation}
where we have used $s_w^2 = 0.2312$. In the following we will often refer to the BSM corrections $\delta c_i \equiv c_i - c_i^{\rm SM}$ ($i = V, A , L ,R$). 

Assuming the Higgs boson is part of an electroweak doublet, the leading corrections are given by the following dimension-$6$ operators
\bea
\label{eq:ttz}
{
\cal L}^{t V}_6& = &\frac{i  c_{Hq}^3}{v^2} H^\dagger \sigma^a D_\mu H \,\bar Q_L \sigma^a \gamma^\mu  Q_L + \mathrm{h.c.} + \frac{i c^1_{Hq}}{v^2} H^\dagger  D_\mu H\, \bar Q_L \gamma^\mu  Q_L + \mathrm{h.c.}\nonumber\\
&&+\frac{i c_{Hu}}{v^2} H^\dagger D_\mu H\,\bar t_R \gamma^\mu  t_R + \mathrm{h.c.}.
\eea
In addition to $c_V$ and $c_A$, these operators affect the $Z b_L b_L$ and $W t_L b_L$ couplings. Deviations of the former from the SM prediction are constrained by LEP data to the per mille level. It is easy to show that this implies, to the same accuracy, the relation $c^1_{Hq} = - c^3_{Hq}$. Then modifications of the $t \bar{t} Z$ interactions are given by
\bea
&&\delta c_V = \frac{1}{4 s_w c_w}\l(2c_{Hq}^3-c_{Hu}\r),\qquad \delta c_A = \frac{1}{4 s_w c_w}\l(-2c_{Hq}^3-c_{Hu}\r).
\eea
On the other hand, tests of the $W t_L b_L$ coupling in single top and $W$ helicity fraction measurements constrain $|c_{Hq}^3|\lesssim 10\%$ (see for example Ref.~\cite{Bernardo:2014vha}).

Direct information on the $Z\bar{t}t$ couplings $c_{V,A}$ can be obtained from the measurement of tree-level processes involving third generation fermions and gauge bosons. With $3$\,ab$^{-1}$ at the $13$\,TeV LHC, the $pp\to t\bar{t}Z$ process can provide determinations of $c_A$ and $c_V$ with relative accuracy of $\sim 0.2$ and $O(1)$, respectively, at $95\%$ CL~\cite{Rontsch:2014cca}. Competitive, and complementary, direct bounds can be derived from the measurement of $tW$ scattering, observable at the LHC in the $pp\to t\bar{t}Wj$ process~\cite{Dror:2015nkp}. 

Notice that, even though in all the models considered in this paper $Z b_L b_L$ is protected at tree level due to the relation $c^1_{Hq} = - c^3_{Hq}$, at $1$-loop the operators in Eq.~\eqref{eq:ttz} generate corrections to the oblique EW observables $S,T$ and to $Z b_L b_L$ itself~\cite{Larios:1999au}, which, if taken at face value, bound their coefficient at the $5\%$ level~\cite{deBlas:2015aea}. Comparable constraints are set by flavor observables~\cite{Brod:2014hsa}. However, since the computation of low-energy observables requires further assumptions (concerning, in particular, the symmetry structure that protects the EW parameters from UV divergences, and the underlying flavor symmetries), a direct measurement of the top-$Z$ couplings remains of the highest priority. 

\begin{figure}
\begin{center}
\includegraphics[scale=0.9]{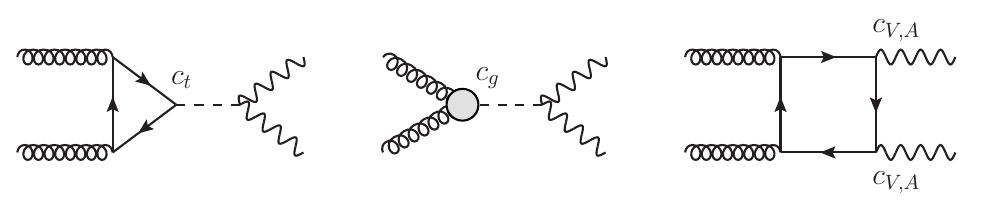}
\caption{Representative subset of the Feynman diagrams for $gg\to ZZ$ that involve the couplings $c_t, c_g, c_V$ and $c_A$. The fermion lines correspond to top quarks.}
\label{fig:feynman}
\end{center}
\end{figure}

Having set up our notation and reviewed the existing bounds, we proceed to the analysis of the $gg\to ZZ\to 4\ell$ process. A sample of the corresponding Feynman diagrams are shown in Fig.~\ref{fig:feynman}, where the $Z$ decays were omitted for simplicity. The predicted number of events in a chosen $\sqrt{\hat{s}}$ bin is in generality a polynomial of the following form
\bea
N&=&a_0+ a_1\, c_A^2+a_2\, c_A^4+ a_{3}\, c_V^2+a_{4}\,c_V^4+ a_{5}\, c_A^2\!\cdot\! c_V^2+
a_6\, c_g + a_7\, c_t + a_8\, c_g^2\nonumber\\
&&+ a_9\, c_t^2 + a_{10}\, c_g\!\cdot\! c_t+
a_{11}\, c_A^2\!\cdot\! c_g+a_{12}\, c_A^2\!\cdot\! c_t  +a_{13}\, c_g\!\cdot\! c_V^2+a_{14}\, c_t\!\cdot\! c_{V}^2\,,\label{eq:interpolation}
\eea
where charge conjugation invariance forbids terms with odd powers of $c_A$ and $c_V$. The numerical coefficients $a_{i}$ were computed using a modified version of MCFM~\cite{Campbell:2010ff,Campbell:2013una} in which the relevant amplitudes are weighted with the couplings $\{c_t, c_g, c_V, c_A\}$, by fitting to a set of simulations performed for various values of the four couplings. The results are presented in Appendix~\ref{sec:simulation}. 

To better understand the constraints in the multi-dimensional coupling space, we compute the standard deviations and correlation matrix after imposing the constraint $c_t+c_g=1$, which we assume will be fixed by on-shell measurements. The result is for the HL-LHC
\begin{figure}
\begin{center}
\includegraphics[scale=0.6]{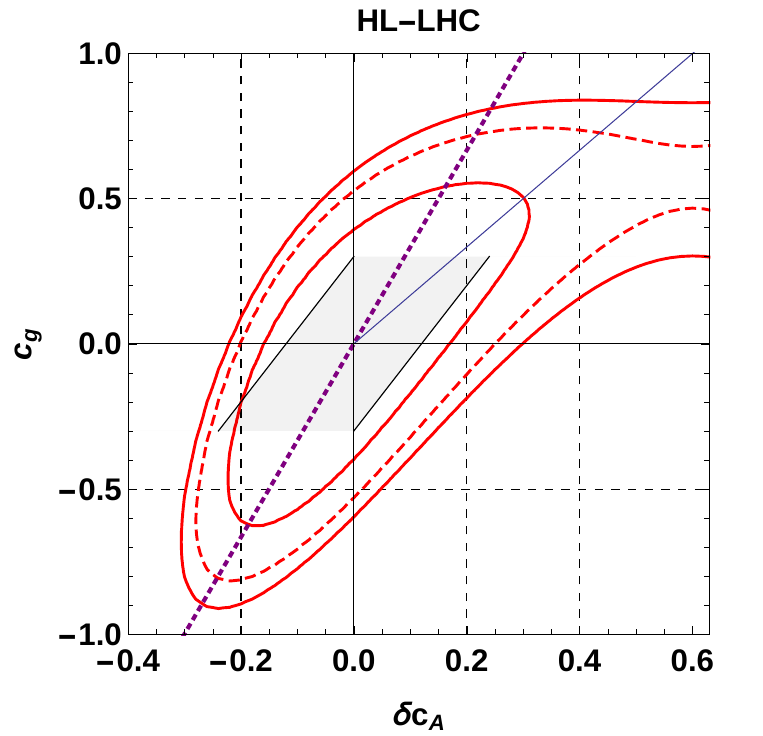}
\includegraphics[scale=0.6]{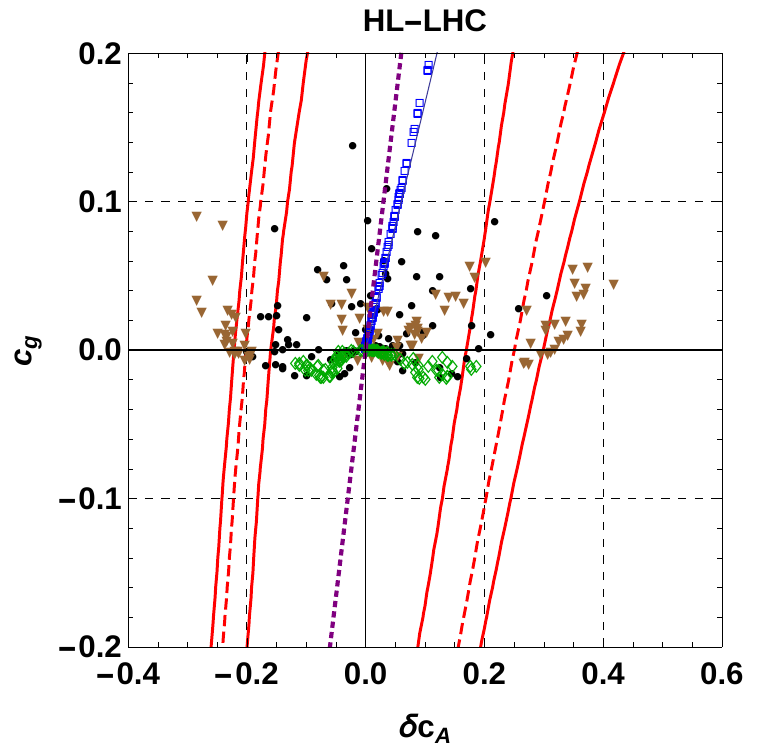}
\includegraphics[scale=0.62]{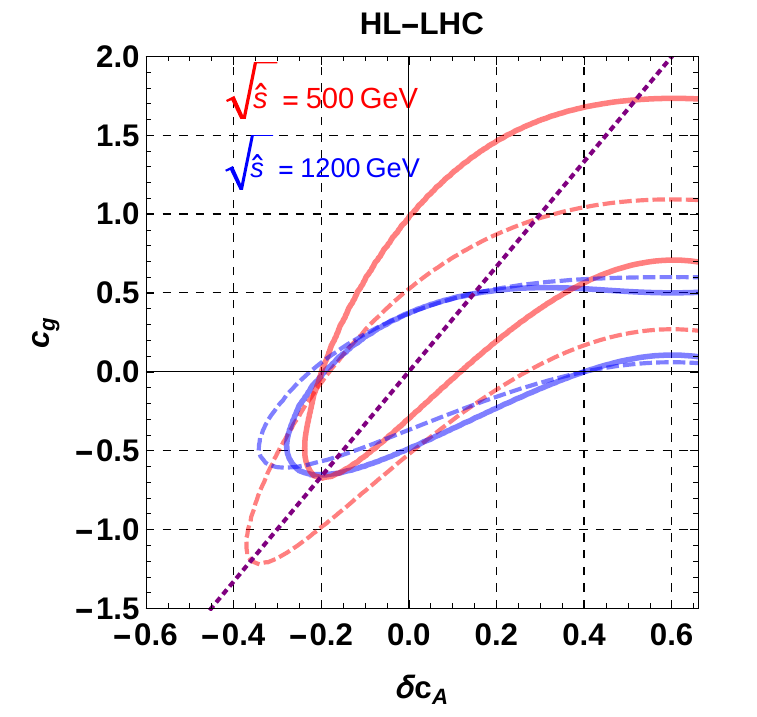}
\includegraphics[scale=0.6]{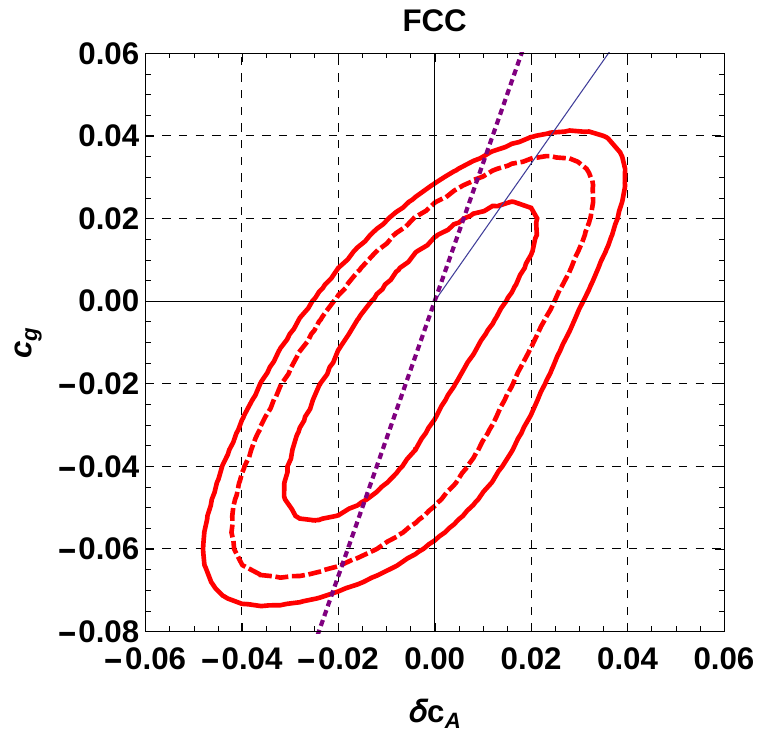}
\end{center}
\caption{\label{fig:corr}{\it Top left panel}: In red, $99,95$ and $68\%$ credibility exclusion contours in the $(\delta c_A, c_g)$ plane from off-shell Higgs measurements at the HL-LHC. The purple dotted line indicates the estimate of the weakest direction of the fit as obtained from a simple analytical expansion, see text for details. The blue line shows the prediction of the singlet top partner model. The grey area is the $95\%$ CL$_s$ constraint from $gg\to hZ$ production as given in Ref.~\cite{Englert:2016hvy}, see Section~\ref{sec:comparison}. {\it  Top right panel}: The same plot, zoomed in near the SM point and overlayed with points showing the predictions of composite Higgs models. Black circles and brown triangles were obtained using the full Lagrangian in Eq.~\eqref{eq:ch}, while green diamonds and blue squares correspond to the predictions of the ${\bf M4_5}$ and ${\bf M1_5}$ simplified models~\cite{DeSimone:2012fs}, respectively. See Section~\ref{sec:eftCH}. {\it Bottom left panel:} The red (blue) solid line shows the $68\%$ credibility contour based on the bin with $\sqrt{\hat s}\in[400,600]\,([1100,1500])$\,GeV. The red (blue) dashed line shows, for illustration, an isocontour of the approximate matrix element squared in Eq.~\eqref{leadingamp}, computed for $\sqrt{\hat s}=500\,(1200)$\,GeV. {\it Bottom right panel:} Exclusion contours in the $(\delta c_A, c_g)$ plane from off-shell Higgs measurements at the FCC. The purple dotted and blue lines indicate the analytical estimate of the weakest direction of the fit and the prediction of the singlet top partner model, respectively.}
\end{figure}
\bea
\l(\baa{c}
 \sigma_{c_A}\\
\sigma_{c_V}\\
 \sigma_{c_g}
\eaa
\r)=\l(\baa{c}
  0.3\\
0.27\\
 0.27
\eaa
\r),\qquad 
\rho=\l(
\baa{ccc}
1&-0.02&0.61
\\
&1&-0.003\\
&&1
\eaa\r),~~
\eea
showing that the strongest correlation is between the parameters $c_A$ and $c_g$. The resulting exclusion contours in the $(\delta c_A, c_g)$ plane\footnote{Notice that the invariance of Eq.~\eqref{eq:interpolation} under $c_A \to - c_A$ translates into a reflection symmetry of the contours around $\delta c_A = - c_A^{\rm SM} \simeq 0.59$, so we restrict to the half of the $(\delta c_A, c_g)$ plane that contains the SM point.} are shown in the top left panel of Fig.~\ref{fig:corr}, where we have set $c_V$ to its SM value (marginalizing over $c_V$ gives a practically identical result, because $c_V$ is very weakly constrained by the fit). We recall that $c_A$ will be tested in the measurements of tree-level processes, such as $pp\to t\bar{t}Z$ and $pp\to t\bar{t}Wj$. In particular, the $pp\to t\bar{t}Z$ analysis of Ref.~\cite{Rontsch:2014cca} finds the $95\%$ CL bound $\delta c_A/c_{A}^{\rm SM} \lesssim 0.2$ with $3$\,ab$^{-1}$ at $13$\,TeV. Interestingly, our results show that the sensitivity of the off-shell Higgs analysis is slightly worse but comparable, thus opening up the opportunity for a competitive test of top-$Z$ interactions in the $gg\to ZZ$ process. This becomes even more relevant once we recall that, differently from our analysis, Ref.~\cite{Rontsch:2014cca} did not include backgrounds. 

It is interesting to investigate further the observed correlation between $c_A$ and $c_g$. We have verified numerically that this behavior is shared by the entire kinematic region with \mbox{$400\; \mathrm{GeV} \lesssim \sqrt{\hat{s}} \lesssim 1.2$\,TeV}. This can be understood thanks to the following simple argument. Let us consider the high energy limit of the leading helicity amplitude $\mathcal{M}^{++00}$, where the $Z$ bosons are longitudinally polarized. Both the top box and top triangle diagrams exhibit a logarithmic divergence at large energy, and the total divergence cancels exactly when the two contributions are weighted with the SM couplings, leaving a UV-finite result. Thus at high energy the leading helicity amplitude has the approximate form
\bea\label{leadingamp}
\mathcal{M}^{++00}(gg\ra ZZ) & \simeq &
 - \, c_g \frac{\hat{s}}{2m_Z^2} + \left(c_t - c_A^2 / c_{A}^{\mathrm{SM}\,2}\right) \frac{m_t^2}{2m_Z^2} \log^2 \frac{\hat{s}}{m_t^2}\nonumber\\
&&-2\pi i \left(c_t - c_A^2 / c_{A}^{\mathrm{SM}\,2}\right) \frac{m_t^2}{2m_Z^2}  \log \frac{\hat{s}}{m_t^2}\,,
\eea
where we have ignored the terms proportional to $\sim c_V^2$, because $(c_V^{\rm SM}/c_A^{\rm SM})^2 \approx  1/7$. For energies $2 m_t \lesssim \sqrt{\hat s} \lesssim 1$\,TeV the imaginary part dominates, therefore we expect the deviation from the SM to be minimized along the direction
\bea
\label{eq:est}
(1-c_g-c_A^2/c^{\mathrm{SM}\,2}_{A})=0 \quad \Rightarrow \quad 
c_g = -\frac{2\delta c_A}{c_{A}^{\mathrm{SM}}} = 8 s_w c_w\, \delta c_A  \simeq 3.4\, \delta c_A\,,
\eea
where the degeneracy condition $c_t = 1 - c_g$ was assumed. As can be read from the top left panel of Fig.~\ref{fig:corr}, this simple estimate of the most weakly constrained direction in the $(\delta c_A ,c_g)$ plane (shown as a dashed purple line) agrees well with the result of the full analysis. Furthermore, in the bottom left panel of Fig.~\ref{fig:corr} we compare the $68\%$ credibility contour obtained restricting the full analysis to $\sqrt{\hat s}\in[400,600]\,([1100,1500])$\,GeV, with an illustrative isocontour of the square of the approximate matrix element in Eq.~\eqref{leadingamp} computed for $\sqrt{\hat{s}} = 500\,(1200)$\,GeV. It is manifest that the exact amplitude squared is qualitatively well approximated by the leading energy terms of Eq.~\eqref{leadingamp}. In addition, the correlation between $\delta c_A$ and $c_g$ is captured by the estimate of Eq.~\eqref{eq:est}, with better accuracy for the bin with lower $\sqrt{\hat{s}}$. At higher energy, $\sqrt{\hat s}\gtrsim 1.2$\,TeV, the real terms of the leading amplitude become more important, and the correlation is altered.

Turning to the FCC analysis, from the bottom right panel of Fig.~\ref{fig:corr} we observe that the expected uncertainty on $c_A$ is roughly $3$-$5\%$ at $1\sigma$. To put this result into context, it is useful to compare it with the expected sensitivity of future $e^+ e^-$ colliders. For $c_A$ recent projections estimate a $\sim 2\%$ uncertainty at the FCC-ee with $\sqrt{s} = 365$\,GeV~\cite{Janot:2015yza}, and $\sim 0.5\%$ at the International Linear Collider with $\sqrt{s} = 500$\,GeV~\cite{Khiem:2015ofa,Amjad:2015mma}. Thus, remarkably, the FCC-hh result is only a factor $2$ weaker than the FCC-ee one. Notice also that the correlation in the $(\delta c_A, c_g)$ plane is not dramatically different from the $14$\,TeV case, indicating that the effect of including the higher-$\sqrt{\hat{s}}$ bins is mild.

\subsection{Implications for composite Higgs models} \label{sec:eftCH}
%
We now turn to discuss the implications of these results for composite Higgs models, the prototypical example of theories where significant corrections to both $c_g$ and $c_{V,A}$ are expected. We begin with the toy model of Eq.~\eqref{eq:singletop}. We find $c_g = 4s_w c_w\, \delta c_A \simeq 1.7\,\delta c_A$, implying that the singlet top partner model is aligned quite closely to the direction of the $(\delta c_A, c_g)$ plane that is most weakly constrained by the off-shell Higgs measurements, as shown in Fig.~\ref{fig:corr}. From the same figure we also infer that models featuring a correlation with opposite sign, {\it i.e.} $c_g\, \delta c_A < 0$, would be subject to much stronger constraints. It is therefore important to investigate the generality of this correlation of sign, by considering more realistic composite Higgs models.  

We focus on models where the Higgs is a pNGB of the spontaneous $SO(5)/SO(4)$ breaking with decay constant $f$. The right-handed top quark is assumed to be a fully composite state arising from the strongly interacting sector, whereas the vector-like top partners $\psi_{1,4}$ transform in the ${\bf 1}$ and $\bf{4}$ representations of $SO(4)$, respectively. Following the notation and conventions of Ref.~\cite{Grojean:2013qca},\footnote{The only departure from the notation of Ref.~\cite{Grojean:2013qca} is the extra tilde on the coefficients of the couplings in the second line of Eq.~\eqref{eq:ch}, which avoids any confusion with the $c_t, c_L$ and $c_R$ defined previously.} the most general Lagrangian is given by
\bea
{\cal L}&=&i\bar
\psi_4 (\sl D+i \sl e)\psi_4 - m_4 \bar{\psi}_4\psi_4 + i \bar \psi_1\sl D\psi_1 - m_1 \bar \psi_1 \psi_1 + i \bar Q_L  \sl{D} Q_L+i \bar t_R\sl Dt_R\nonumber\\
&&+ i \tilde{c}_t \bar \psi_{4 R}^{i} \sl d^{\,i} t_R + i \tilde{c}_R \bar \psi_{4 R}^i \sl d^{\,i} \psi_{1R}+ i \tilde{c}_L \bar \psi_{4 L}^i \sl d^{\,i} \psi_{1 L} + \mathrm{h.c.} \nonumber \\
&&+ y_{Lt} f (\bar Q_L)^IU_{I5}t_{R} + y_{L4} f (\bar Q_L)^IU_{Ii}\psi_{4 R}^i
+y_{L1} f (\bar Q_L)^IU_{I5}\psi_{1 R} + \mathrm{h.c.},
\label{eq:ch}
\eea
which in the limit $m_{1}\,(m_4) \ra \infty$ reduces to that of the ${\bf M4_5\,(M1_5)}$ model studied in Ref.~\cite{DeSimone:2012fs}. The $hgg$ effective coupling and the corrections to the $Z\bar{t} t$  interactions are given, at first order in $v^2$, by\footnote{The last two formulas in Eq.~\eqref{eq:chtt} were already given in Ref.~\cite{Grojean:2013qca}.}
\begin{eqnarray}
&\displaystyle c_g \,=\, \frac{v^2}{2}\l(\frac{y_{L1}^2 m_4^2}{m_1^2(m_4^2 + y_{L4}^2 f^2)} - \frac{y_{L4}^2}{m_4^2 + y_{L4}^2 f^2 } +\frac{y_{L4}^2 y_{Lt}^2 f^2 }{(m_4^2 + y_{L4}^2 f^2)^2}\r),
\nonumber  \\
\label{eq:chtt}
&\displaystyle \left(\frac{1}{2} - \frac{2}{3}s_w^2\right) \delta c_L \,=\, - \frac{v^2}{4} \frac{(y_{L4}^2 m_1^2 + y_{L1}^2 m_4^2 - 2 \sqrt{2} \tilde{c}_L y_{L4} y_{L1} m_1 m_4 )}{m_1^2(m_4^2 + y_{L4}^2 f^2) }\,, \nonumber
\\
&\displaystyle \left( - \frac{2}{3}s_w^2\right) \delta c_R \,=\, \frac{v^2}{4} \frac{( y_{L4}^2 y_{Lt}^2 f^2 - 2 \sqrt 2 \tilde{c}_t y_{L4} y_{Lt} (m_4^2 + y_{L4}^2 f^2))}{(m_4^2 + y_{L4}^2 f^2)^2}\,,
\nonumber 
\end{eqnarray}
where $\delta c_{L,R}$ are related to the corrections to the vector and axial couplings $c_{V,A}$ by
\begin{equation}
\delta c_{V,A} = \frac{1}{2e}\left(\pm \delta c_L\, g_L^{\rm SM} + \delta c_R\, g_R^{\rm SM}\right).
\end{equation}
We see that, in general, the signs of $c_g$ and $\delta c_A$ are not correlated. To illustrate this point we performed a numerical scan, whose results are presented in the top right panel of Fig.~\ref{fig:corr}. We set $f=800$\,GeV, while the composite fermion masses $m_{1,4}$ were varied in the range $[0.8,1.5]$\,TeV. The coefficients of the derivative interactions were fixed to $\tilde{c}_t = \tilde{c}_L = 3$. The black points correspond to values of the Yukawa couplings $y_{Lj} \in [0,2]$ ($j=t, 1, 4$) and the brown points to $y_{Lj} \in [2,3]$. Blue points sitting almost exactly on the singlet partner line are the predictions of the ${\bf M1_5}$ model, whereas the green points are the predictions of the ${\bf M4_5}$ model; in both cases, the Yukawa couplings were varied in the interval $[0,3]$. The results show that in a sizable fraction of the parameter space of Eq.~\eqref{eq:ch}, the correlation between $c_g$ and $\delta c_A$ has opposite sign compared to the toy model. Thus off-shell Higgs production can set significant constraints on composite Higgs models.

As a concluding remark, it is worthwhile to comment on the sign of the $Z \bar{t}t$ corrections when the derivative interactions in the second line of Eq.~\eqref{eq:ch} are turned off. In this case we find $\delta c_L < 0$, which can be understood with the following observation. In order to protect the $Z b_L b_L$ coupling, the $Q_L$ doublet must be embedded in the $\mathbf{4}_{2/3}$ representation of the custodial $O(4) \times U(1)_X$ symmetry~\cite{Agashe:2006at}. This leaves only two possible choices for the embedding of $t_R$, $\mathbf{1}_{2/3}$ or $\mathbf{6}_{2/3}$. In both cases, the $t_L$ can mix only with vector-like fermions that have $\left|T_{L}^3\right| \leq 1/2$, leading to $\delta c_L < 0$. On the other hand, from Eq.~\eqref{eq:chtt} we also read that $\delta c_R < 0$, but this result is more model-dependent: for example, if the $t_R$ is only partially composite one finds a small, positive $\delta c_R$~\cite{Grojean:2013qca}. Based on these results, we expect that in general $\delta c_A > 0$ will be preferred. It is however important to stress that when the derivative interactions parameterized by the coefficients $\tilde{c}_i$ are present, either sign is possible.

\subsection{Comparison to other gluon-fusion processes} \label{sec:comparison}
As anticipated in Section~\ref{sec:projections}, we now return to the $gg\to hZ$ process. As first pointed out in Ref.~\cite{Bylund:2016phk} and further studied in Ref.~\cite{Englert:2016hvy}, its amplitude is sensitive, in addition to $c_y$, to modifications of the $Z\bar{t}t$ interactions. Since at high energy both $gg\to ZZ$ and $gg\to hZ$ are dominated by loops of top quarks, we can gain some understanding on the expected sensitivity of the two processes to the top-$Z$ couplings by inspecting the tree-level scatterings $t\bar{t}\to ZZ, hZ$ at large $\sqrt{\hat{s}}$. These can be obtained from 
the relative loop diagrams by means of $s$-channel cuts. In the presence of the operators of Eq.~\eqref{eq:ttz}, $t\bar{t}\to hZ$ is dominated by the interaction with schematic form $h 
\partial_\mu \chi_a (\bar{\psi}\gamma^\mu \psi)_a/v^2$ (where $\psi \in \{t,b\}$, $\chi_a$ are the Goldstone bosons eaten by the $W$ and the $Z$, and $h$ is the physical Higgs boson), 
which leads to a strong growth of the amplitude $\sim \hat{s}/v^2$. For $t\bar{t}\to ZZ$, the corresponding leading interaction is $\epsilon_{abc}\chi_b \partial_\mu \chi_c 
(\bar{\psi}\gamma^\mu \psi)_a/v^2$, which however vanishes when two longitudinal $Z$'s are selected. As a consequence, the amplitude for $t\bar{t}\to ZZ$ only grows as $m_t 
\sqrt{\hat{s}}/v^2$~\cite{Dror:2015nkp}. This simple observation hints that $gg\to hZ$ should have a stronger sensitivity to $t\bar{t}Z$ modifications than $gg\to ZZ$. Indeed, reinterpreting the results of Ref.~\cite{Englert:2016hvy} we find\footnote{We make use of $\delta c_A = - \bar{c}_{Ht}/(4s_w c_w)$ and $c_g = - \bar{c}_t$, where the barred coefficients were defined in Ref.~\cite{Englert:2016hvy} and to obtain the second relation we have assumed the degeneracy condition $c_g = c_y$.} that the constraint on the $(\delta c_A, c_g)$ plane obtained from $gg\to hZ$ is somewhat stronger than the one from our $gg\to ZZ$ analysis, see the top left panel of Fig.~\ref{fig:corr}. We also observe that, interestingly, the two constraints are approximately aligned, although the origin of this alignment is not transparent. Finally, the above analysis of tree-level subamplitudes also points to $gg\to WW$~\cite{Duhrssen:2005bz} as a promising process to constrain top-$Z$ coupling modifications, since the $\chi^+ \chi^-\bar{t}t$ vertex is generated by the corresponding dimension-$6$ operators. A detailed study of the $WW$ channel would be an interesting extension of this paper.

\section{Validity of the EFT for off-shell Higgs}
\label{sec:EFTvalidity}

In the previous section we have shown that a general EFT treatment of off-shell Higgs production must go beyond the two operators in Eq.~\eqref{eq:L6}, by including also operators that modify the still weakly constrained top-$Z$ interactions. In this section, instead, we focus on testing the validity of the EFT as a description of the low-energy effects of the underlying BSM physics. We achieve this by comparing the EFT and exact predictions for the toy model of Eq.~\eqref{eq:singletop}.  

Before presenting our quantitative results, it is useful to recall the parametric conditions that need to be satisfied for the EFT description to be valid~\cite{Contino:2016jqw}:
\bit
\item Small energy requirement: the EFT is valid only at energies $E$ below the masses of the new resonances,
\bea
\label{eq:eftv}
\frac{E}{M_*}\ll 1 \;;
\eea
\item Small coupling requirement: since every insertion of the Higgs boson is accompanied by the coupling $Y_*$, the EFT expansion is valid only if
\bea
\label{eq:couplexp}
\frac{Y_* v}{M_*}\ll 1\;;
\eea
\item Suppression of dimension-$8$ operators: since in our study we include only dimension-$6$ operators, we need to require that the contribution of operators of higher dimension be subleading. The dimension-$8$ effects can be parameterized, for example, by
\begin{equation}
O_g^{(8)} \sim \frac{g_s^2}{16\pi^2}\frac{Y_\ast^2}{M_\ast^4}|D_{\lambda}H|^2 G_{\mu\nu}G^{\mu\nu}\,.
\end{equation}
Comparing with Eq.~\eqref{eq:opH}, we find that $O_g^{(8)}$ is subleading to $O_g$ for $E \ll M_\ast$~\cite{Azatov:2014jga}, {\it i.e.} in this model the dimension-$8$ effects are automatically suppressed.\footnote{Notice, however, that this is not true in more realistic composite Higgs models, where the Goldstone nature of the Higgs gives an extra suppression of $O_g$. In this case the dimension-$8$ effects can only be neglected for $E \ll y_t f$~\cite{Azatov:2015oxa}.}
\eit

\begin{figure}
\begin{center}
\includegraphics[scale=0.9]{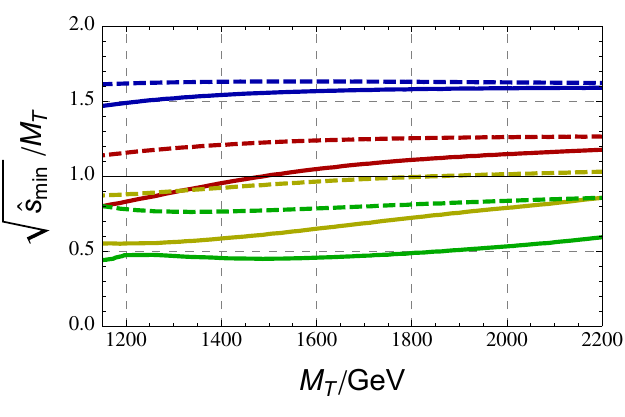}
\end{center}
\caption{\label{eft} The ratio of the minimal partonic energy for which the EFT description becomes invalid to the mass $M_T$ of the vector-like top partner, as function of $M_T$. Several values of the Yukawa coupling $Y_*$ are considered: $Y_* = 2,3,4,5$ are indicated by the blue, red, yellow and green curves, respectively. The solid lines correspond to the dimension-$6$ EFT, whereas the dashed lines correspond to the nonlinear parameterization where the full modifications to the $h \bar{t}t, Z \bar{t}t$ and $hgg$ couplings are retained.} 
\end{figure}

We now proceed to an explicit comparison between the EFT and the exact prediction of the toy model. The latter was computed using the FeynArts/FormCalc/LoopTools combination~\cite{Hahn:2000kx,Hahn:1998yk}. Given the partonic differential cross section $d\hat{\sigma}/d\hat{s}$, we can define the region of validity of the EFT description as
\bea 
\label{eq:EFTbreakdown}
\frac{\left| \left( \frac{d \hat{\sigma}}{d\hat{s}} \right)_{\hbox{\scriptsize full}}-\left( \frac{d \hat{\sigma}}{d\hat{s}} \right)_{\hbox{\scriptsize EFT}}\right|}{\left(\frac{d \hat{\sigma}}{d\hat{s}}\right)_{\hbox{\scriptsize full}}}< 0.05\,.
\eea
The minimal energy $\sqrt{\hat{s}_{\rm min}}$ for which Eq.~\eqref{eq:EFTbreakdown} is not satisfied is shown in Fig.~\ref{eft} as function of the physical top partner mass $M_T$, for some choices of the coupling $Y_*$. In addition to the dimension-$6$ EFT based on Eq.~\eqref{eq:opH}, we consider an approximation where  the effective couplings in Eqs.~(\ref{eq:tpEFTtree}\,,\ref{eq:EFTloop}) are computed at all orders in $1/M_\ast$, which we label `nonlinear parameterization.' As expected, we find that the EFT approximation breaks down at energies close to the resonance mass, $\sqrt{\hat{s}} \sim M_*$. In addition, the nonlinear parameterization gives a better approximation to the full theory compared to the EFT. This effect is more noticeable for larger $Y_\ast$, because the nonlinear parameterization includes the resummation of the terms of higher order in $(Y_* v/M_*)^2$, which are neglected in the EFT.

%
\begin{figure}
\begin{center}
\includegraphics[scale=0.55]{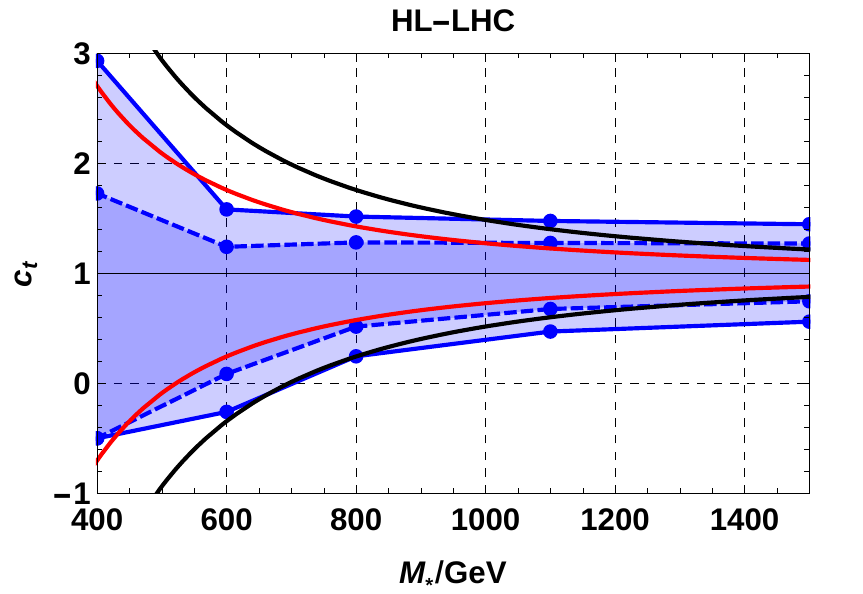}
\includegraphics[scale=0.55]{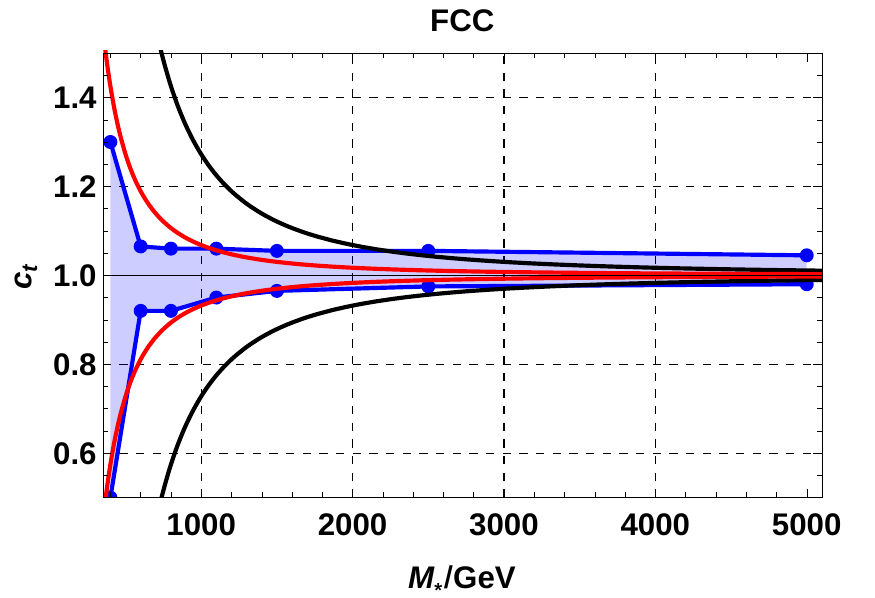}
\end{center}
\caption{\label{fig:eftL14} {\it Left panel:} The blue solid (dashed) line indicates the $95\%$ (68$\%$) HL-LHC bound on $c_t$ as a function of the maximum energy included in the analysis, $M_\ast$. The red (black) lines show the expected parametric dependence of $c_t$ on the mass $M_*$ of a resonance, $c_t = 1 + a Y_*^2 v^2/M_*^2$ with $a = \pm 1$ and $Y_\ast = 3\,(4)$. {\it Right panel:} Same as left panel, for the FCC analysis. The red (black) lines correspond to $a = \pm 1$ and $Y_\ast = 1.5\,(3)$.}
\end{figure}

We have seen that below $\sqrt{\hat{s}}\sim M_*$, the EFT provides an accurate description of the underlying UV theory. Then one can ask how much the constraints on the effective coefficients degrade, if the categories with higher energy are removed from the analysis. To address this question, we have performed a simplified analysis where modifications of the top-$Z$ couplings are neglected and the degeneracy condition $c_t+c_g=1$ is assumed, leaving only $c_t$ as free parameter. The results are presented in Fig.~\ref{fig:eftL14}, where the constraint on $c_t$ is shown as a function of the maximum energy of the events kept in the analysis, labeled $M_\ast$. For illustration purposes, in the same figure we have also drawn the contours showing the expected parametric dependence of $c_t$ on the mass $M_\ast$ of a resonance, $c_t = 1 + a Y_*^2 v^2/M_*^2$ with $a \sim O(1)$, for some representative values of the coupling $Y_\ast$. We take $Y_*=3,\,4$ for the HL-LHC analysis and $Y_* = 1.5,\,3$ for the FCC, and $a = \pm 1$. For example, for $a= -1$ and $Y_\ast = 3$ we find that the $95\%$ CL bound obtained from the HL-LHC analysis including all bins would read $M_\ast \gtrsim 800$\,GeV, whereas removing the events with energy above $M_\ast$ gives $M_\ast \gtrsim 400$\,GeV. Similary, for $a = -1$ and $Y_\ast = 1.5$ the full FCC analysis would yield $M_\ast \gtrsim 1.6$\,TeV, but after removing the high-energy bins we find that only the region $500\;\mathrm{GeV}\lesssim M_\ast \lesssim 1.2$\,TeV is actually excluded. These results stress the importance of a consistent EFT treatment to avoid over-estimating the exclusion bounds.  

\begin{figure}
\begin{center}
\includegraphics[scale=0.5]{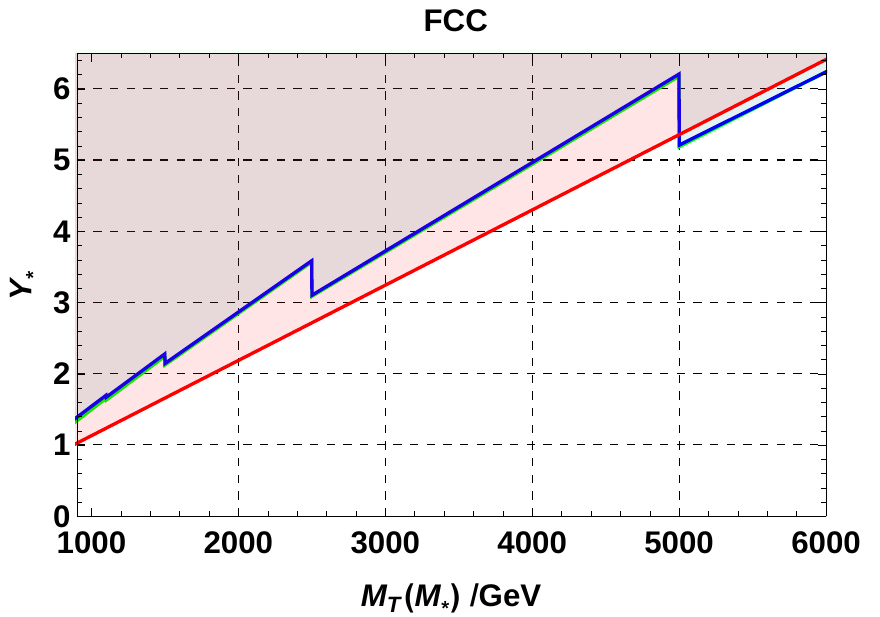}
\includegraphics[scale=0.5]{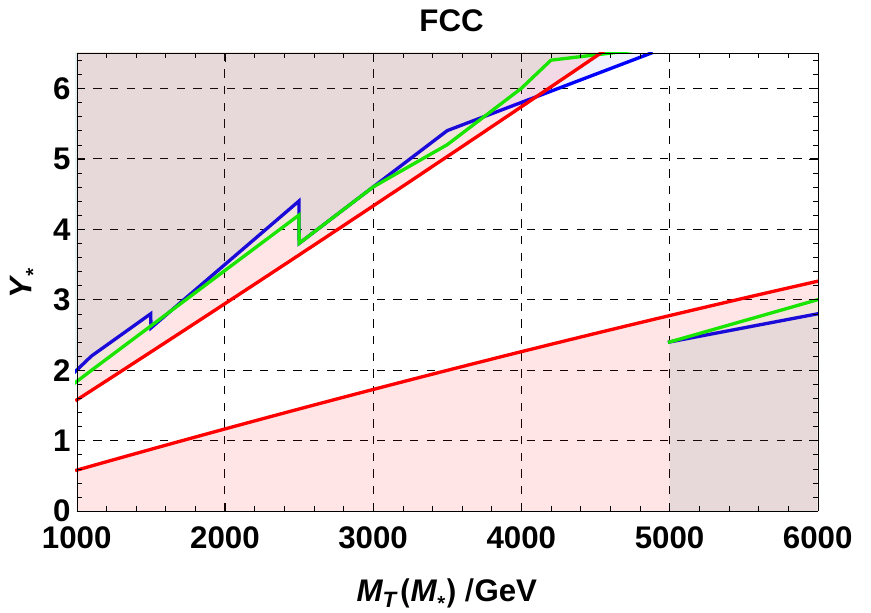}
\end{center}
\caption{\label{fig:ymFCC} {\it Left panel:} $2\,\sigma$ constraints on the singlet top partner model at the FCC, assuming the observed number of $4\ell$ events to agree with the SM prediction. The red region is obtained using the full model simulation, while the blue and green regions correspond to the EFT and nonlinear parameterization, respectively. In the EFT case (blue), the horizontal axis shows $M_*$ defined in Eq.~\eqref{eq:singletop} instead of the exact top partner mass $M_T$. {\it Right panel:} Same as in the left panel, but assuming the observed number of $4\ell$ events to equal the prediction of the singlet top partner model with $M_{T}=3$\,TeV and $Y_*=3.5$.}
\end{figure}

At last, we compare the bounds obtained from a full computation in the toy model of Eq.~\eqref{eq:singletop} with those from the EFT analysis. At the HL-LHC the constraints turn out to be very weak, 
because the toy model lies approximately along the least constrained direction in the $(\delta c_A,\,c_g)$ plane (see the discussion in Section~\ref{sec:ttZeffects}), so we proceed directly to 
the FCC predictions. The results are presented in Fig.~\ref{fig:ymFCC}. In the left panel we have assumed the observed number of $4\ell$ events to agree with the SM prediction. The area shaded in red is the exclusion derived from the full calculation, while the blue and green regions are the exclusions obtained using the EFT and the nonlinear parameterization, respectively. In the last 
two analyses only the bins with $\sqrt{\hat{s}}$ below the mass of the hypothetical resonance are kept, leading to the `spiky' shape of the bounds. For small values of the top 
partner mass $M_T$, the full calculation gives a stronger constraint because it retains the tail of the invariant mass distribution, which is discarded in the EFT and nonlinear analyses. On 
the other hand, since we always neglect events with $\sqrt{\hat{s}}$ above $5$\,TeV, for $M_T$ larger than this value the discrepancy between the EFT/nonlinear and the full 
calculation decreases. In this high mass region, the only difference between the EFT and full treatments is given by operators with dimension $>6$, which are neglected in the EFT, 
whereas the difference between the nonlinear parameterization and the full computation arises from operators with more than two derivatives, whose effects are not captured by the 
nonlinear analysis. 

In the right panel of Fig.~\ref{fig:ymFCC} we have instead assumed that a BSM signal, given by the singlet top partner model with $M_{T}=3$\,TeV and $Y_*=3.5$, will be observed at the FCC. In this case, both the EFT/nonlinear analyses and the one based on the full calculation would be able to reject the SM hypothesis $(M_T\ra \infty,\,Y_*\ra 0)$ at the $2\,\sigma$ level. Interestingly, however, the full analysis can set a non-vanishing lower bound on $Y_\ast$ in the whole range of hypothetical resonance masses, whereas the EFT/nonlinear analyses are able to achieve this only for masses above $5$\,TeV. This is due to the important effect of the last invariant mass bin with $\sqrt{\hat{s}}\in[2.5,5]$\,TeV, which in the EFT and nonlinear analyses is included only for $M_T> 5$\,TeV.

\section{Summary and outlook} \label{sec:summary}

The main target of this paper was the well-known degeneracy that does not allow LHC inclusive Higgs measurements to disentangle BSM corrections to the $h\bar{t}t$ coupling, parameterized by the dimension-$6$ operator $\sim c_y |H|^2 \bar{Q}_L \widetilde{H} t_R$, from contributions to the contact operator $\sim c_g |H|^2 G_{\mu\nu}G^{\mu\nu}$. Processes that have been proposed to resolve this degeneracy include $t\bar{t}h$, boosted, off-shell and double Higgs production. In Section~\ref{sec:projections} we have presented, building on previous results available in the literature and employing an EFT based on $c_y$ and $c_g$, HL-LHC and FCC projections for all these measurements.

We then proceeded to take a critical look at the applicability of the EFT approach, focusing on off-shell Higgs production. We started by questioning whether other operators beyond $O_y$ and $O_g$ can impact our results, finding that corrections to the top-$Z$ couplings, being weakly constrained by LHC current and future direct measurements, can alter the box diagram contribution to $gg\to ZZ$ at a significant level. Furthermore, in composite Higgs models the corresponding operators are typically generated with coefficients of size comparable to that of $c_y$ and $c_g$, as we illustrated using a simplified example with a single vector-like quark. This motivated us to perform in Section~\ref{sec:ttZeffects} an extended EFT analysis of off-shell Higgs production, where generic modifications of the top-$Z$ interactions were included. We found that the SM unitarity preservation at high energy forces a very strong correlation between modifications of the top-$Z$ and top-Higgs interactions, and leads to a weakly constrained direction in the coupling space. Interestingly, our toy model, as well as more realistic composite Higgs models with a light singlet top partner, sit approximately along this direction. By performing a more general analysis, however, we showed that other realizations do not share this feature. 

Interestingly, our analysis showed that despite being loop suppressed, $gg\to ZZ$ can compete with tree-level processes, such as $t\bar{t}Z$ or $t\bar{t}Wj$ production, in constraining corrections to the top-$Z$ couplings. Furthermore we pointed out that, due to the symmetry structure of the relevant dimension-$6$ operators, $gg\to WW$ may be even more effective than $gg\to ZZ$ for this purpose. This warrants a dedicated study of the $WW$ process at high invariant masses, including the relevant backgrounds.

The power of off-shell (as well as of boosted and double) Higgs production to discriminate between $c_y$ and $c_g$ is a consequence of probing the kinematic regions where $\sqrt{\hat{s}}\gg m_t$. This can lead to concerns about the validity of the EFT treatment. In Section~\ref{sec:EFTvalidity} we analyzed this point in detail. By using a toy model with a single top partner, we explicitly verified the range of applicability of the EFT, finding agreement with the bottom-up estimate. We finally compared the bounds obtained from a full calculation to those derived within the EFT, stressing the dependence of the latter on the largest energy scale considered in the analysis. 

To conclude, we believe that the results of this paper constitute significant progress towards a global, consistent EFT analysis of Higgs and top data at hadron colliders, from which the first clues to the solution of the naturalness puzzle may come to light.


\vspace{8mm}
\noindent {\bf Note added:} While this project was being completed, Ref.~\cite{Maltoni:2016yxb} appeared whose results partially overlap with those presented in Section~\ref{sec:projections} of this paper. In particular, the degeneracy between $c_y$ and $c_g$ was also addressed in Ref.~\cite{Maltoni:2016yxb}, by combining the measurements of inclusive, $t\bar{t}$-associated and boosted Higgs productions at the HL-LHC. We find agreement with that projection. Our analysis differs from that of Ref.~\cite{Maltoni:2016yxb} in several aspects: Here the roles of double and off-shell Higgs productions in resolving the degeneracy were also investigated, and the Higgs decays were included. In addition, we presented projections for the FCC. On the other hand, we neglected the effects of the chromo-magnetic top dipole operator, which were extensively studied in Ref.~\cite{Maltoni:2016yxb}.

\vspace{8mm}

\noindent{\bf Acknowledgments}  We are grateful to R.~Contino, G.~Panico and M.~Son for sharing with us their results on double Higgs production, and to M.~Schlaffer, M.~Spannowsky and A.~Weiler for detailed explanations on their boosted Higgs analysis. We thank G.~Cacciapaglia, A.~Deandrea, G.~Drieu La Rochelle and J.~B.~Flament for collaboration in the early stage of this project. We are grateful to M.~Luty for useful discussions, and to F.~Maltoni, E.~Vryonidou, C.~Zhang and J.~Zupan for comments about the first version of the manuscript. E.~S. thanks J.~Dror, M.~Farina and J.~Serra for many discussions on topics related to this paper. C.~G. is supported by the European Commission through the Marie Curie Career Integration Grant 631962 and by the Helmholtz Association. A.~P. is supported by the European Research Council under the European Union's Seventh Framework Programme (FP/2007-2013)/ERC Grant Agreement n. 279972 `NPFlavour'. E.~S. is supported in part by the US Department of Energy grant DE-SC-000999, and thanks the MIAPP for hospitality and partial support. A.~A., C.~G. and A.~P. thank the Centro de Ciencias de Benasque Pedro Pascual for its hospitality while this project was being developed. All authors acknowledge hospitality from the GGI, where part of this work was done, and partial support from the INFN.

\vspace{4mm}
\section*{Appendices}
\vspace{2mm}
\appendix
\section{Off-shell Higgs analysis}
\label{sec:simulation}
In this appendix we summarize the results of our off-shell Higgs analysis. For further details, we refer the reader to Ref.~\cite{Azatov:2014jga}. We generated the process $gg \to ZZ \to 4\ell$ using MCFM v6.8~\cite{Campbell:2010ff,Campbell:2013una}, modified for the effective couplings. The result was cross-checked against an independent FeynArts/FormCalc/LoopTools~\cite{Hahn:2000kx,Hahn:1998yk} implementation. The generation was performed at the leading order in QCD, and scaled to NLO by applying an invariant mass-dependent $K$-factor~\cite{Ball:2013bra,Azatov:2014jga}. Notice that recently, important progress was made toward a full NLO computation of $gg\to ZZ$, by applying a large-$m_t$ expansion to the only piece that still remains exactly unknown, the two-loop continuum production through top loops~\cite{Melnikov:2015laa,Campbell:2016ivq,Caola:2016trd}. In particular, Ref.~\cite{Campbell:2016ivq} found that at the $13$\,TeV LHC the $K$-factors for the Higgs amplitude squared and for the Higgs-continuum interference agree within $5\%$ in the region $\sqrt{\hat{s}}> 250$\,GeV, which we consider here. This supports the prescription proposed in Ref.~\cite{Bonvini:2013jha} and adopted in Ref.~\cite{Azatov:2014jga}, consisting in applying a single, invariant mass-dependent $K$-factor to the entire $gg\to ZZ$ amplitude squared, which we have maintained in this paper. Finally, the non-interfering background $q\bar{q}\to ZZ \to 4\ell$ was simulated in MCFM at NLO. The MSTW2008 parton distribution functions (PDFs)~\cite{Martin:2009iq} were used.
\subsection{14\,TeV}

For the $14$\,TeV analysis we bin the $4\ell$ invariant mass distribution as follows,
\bea
\sqrt{\hat{s}}=(250,400,600,800,1100,1500)\;\mathrm{GeV}.
\eea
The corresponding $gg\to 4\ell$ yields are, for $1$\,ab$^{-1}$, 
\begin{eqnarray}
N^{14}_{[250,400]} &=& 
-173\, c_A^2\!\cdot\! c_g - 266\, c_A^2\!\cdot\! c_t + 8.51\, c_A^2\!\cdot\! c_V^2 + 185\, c_A^4 - 0.749\, c_A^2
\nonumber\\
&&
+181\,  c_g\!\cdot\! c_t+1.95\, c_g\!\cdot\! c_V^2 + 63.9\, c_g^2 - 104\, c_g + 5.06\, c_t\!\cdot\! c_V^2\nonumber\\
 && + 132\, c_t^2 - 138\, c_t + 10.8\, c_V^4 - 124\, c_V^2 + 2300\,,
\nonumber\\
N^{14}_{[400,600]} &=& 
-175\, c_A^2\!\cdot\! c_g - 452\, c_A^2\!\cdot\! c_t + 9.19\, c_A^2\!\cdot\! c_V^2 + 463\, c_A^4 + 45.9\, c_A^2
\nonumber\\
&&
+130\, c_g\!\cdot\! c_t + 1.09\, c_g\!\cdot\! c_V^2 + 48.0\, c_g^2 - 12.9\, c_g + 3.11\, c_t\!\cdot\! c_V^2
   \nonumber\\
&&
+140\, c_t^2 - 22.9\, c_t + 8.27\, c_V^4 - 3.93\, c_V^2 + 294\,,
\nonumber\\
 N^{14}_{[600,800]} &=&
-33.1\, c_A^2\!\cdot\! c_g - 188\, c_A^2\!\cdot\! c_t + 2.24\, c_A^2\!\cdot\! c_V^2 + 235\, c_A^4 + 10.7\, c_A^2\nonumber\\
&&
+31.8\, c_g \!\cdot\!  c_t - 0.271\, c_g\!\cdot\! c_V^2 + 27.0\, c_g^2 - 1.48\, c_g + 0.278\, c_t\!\cdot\! c_V^2
 \nonumber\\
&&
   +46.0\, c_t^2 -1.44\, c_t + 1.68\, c_V^4 + 11.4\, c_V^2 +37.0 \,,
\nonumber\\
 N^{14}_{[800,1100]} &=&
 4.07\, c_A^2\!\cdot\! c_g - 90.5\, c_A^2\!\cdot\! c_t + 0.796\, c_A^2\!\cdot\! c_V^2 + 124\, c_A^4 + 3.25 c_A^2
    \nonumber\\
&&
 +7.42\, c_g\!\cdot\!  c_t - 0.204\, c_g\!\cdot\! c_V^2 + 21.6\, c_g^2 - 0.259\, c_g + 0.0960\, c_t \!\cdot\! c_V^2
      \nonumber\\
&&
+19.\,3 c_t^2 - 0.127\, c_t + 0.647\, c_V^4 + 4.49\, c_V^2 + 8.78\,,
\nonumber\\
 N^{14}_{[1100,1500]} &=&
10.4\, c_A^2\!\cdot\! c_g - 28.4\, c_A^2\!\cdot\! c_t + 0.127\, c_A^2\!\cdot\! c_V^2 + 41.0\, c_A^4 + 0.891\, c_A^2
\nonumber\\
&&
 - 0.783\,c_g\!\cdot\! c_t - 0.0263\, c_g\!\cdot\! c_V^2 + 13.1\, c_g^2 - 0 .0195\, c_g + 0.0876\, c_t\!\cdot\! c_V^2
         \nonumber\\
&&
   +5.50\, c_t^2 - 0.052\, c_t + 0.151\, c_V^4 + 1.02\, c_V^2 + 1.58
 \,.
 \label{eq:yield}
\end{eqnarray}
These numbers were obtained by assuming the identification efficiency for each lepton is $95\%$, summing over all the charge/flavor final states and applying the following $K$-factors for each bin~\cite{Azatov:2014jga,Ball:2013bra}
\bea
K=\{1.96,1.86,1.81,1.80,1.81\}.
\eea
The $q\bar{q}\to 4\ell$ background yields\footnote{Notice that, due to a numerical mistake, in Eq.~(3.20) of Ref.~\cite{Azatov:2014jga} we reported background yields that were $\sim 5\%$ larger than the correct ones, which appear in Eq.~\eqref{qqbar_14}. The effect on the results of Ref.~\cite{Azatov:2014jga} is negligible.} are, for $1$\,ab$^{-1}$,
\begin{equation} \label{qqbar_14}
N^{14}_{q\bar{q}} = \{10100, 2220, 450, 164, 44.5\}.
\end{equation}

\subsection{100\,TeV}
 
The $100$\,TeV analysis is very similar to the $14$\,TeV one, but includes events with $4\ell$ invariant mass up to $5$\,TeV, with the binning
\begin{equation}
\sqrt{s} = (250, 400, 600, 800, 1100, 1500, 2500, 5000)\;\mathrm{GeV}. 
\end{equation}
In principle, the analysis could be extended to even higher invariant masses. However, the cross section drops off fast with $\sqrt{\hat{s}}$, hence the simulation time increases correspondingly. In particular, simulations involving $c_V$, which is weakly constrained by the fit, become a potential issue at very high $\sqrt{\hat{s}}$. As a result we chose to restrict our analysis to $5$\,TeV. The $gg\to 4\ell$ yields are, for $1$\,ab$^{-1}$, 
\begin{eqnarray}
\label{eq:yieldFCC}
N^{100}_{[250,400]}&=&-2950\, c_A^2\!\cdot\! c_g-4540\, c_A^2\!\cdot\! c_t+171\, c_A^2\!\cdot\! c_V^2+3180\, c_A^4-36.8\, c_A^2
\nonumber\\
&&
+3130\, c_g\!\cdot\!c_t+60.0\, c_g\!\cdot\! c_V^2+1110\, c_g^2 -1810\, c_g  + 95.7\, c_t\!\cdot\! c_V^2
      \nonumber\\
&&+2240\, c_t^2-2320\, c_t+153\, c_V^4 - 2170\, c_V^2 + 39400\,,
\nonumber\\
N^{100}_{[400,600]} &=&
-4530\, c_A^2\!\cdot\! c_g - 11800\, c_A^2\!\cdot\! c_t + 229\, c_A^2\!\cdot\! c_V^2 + 12100\, c_A^4 + 1170\, c_A^2
\nonumber\\
&&
+3360\, c_g\!\cdot\!  c_t + 19.5\, c_g\!\cdot\! c_V^2 + 1250\, c_g^2 - 326\, c_g + 88.8\, c_t\!\cdot\! c_V^2
\nonumber\\
&&
   +3610\, c_t^2 - 571\,  c_t + 225\, c_V^4 - 111\, c_V^2 + 7360\,,
\nonumber\\
N^{100}_{[600,800]}&=&
-1280\, c_A^2\!\cdot\! c_g - 7240\, c_A^2\!\cdot\! c_t + 87.1\, c_A^2\!\cdot\! c_V^2 + 9080\, c_A^4 + 418\, c_A^2
\nonumber\\
&&
+1220\, c_g\!\cdot\!   c_t - 8.13\, c_g\!\cdot\! c_V^2 + 1040\, c_g^2 - 53.7\, c_g + 16.8\, c_t\!\cdot\! c_V^2
\nonumber\\
&&
   +1780\, c_t^2 - 87.0\, c_t + 82.2\, c_V^4 + 407\, c_V^2 + 1380\,,
\nonumber\\
N^{100}_{[800,1100]}&=&
265.\, c_A^2\!\cdot\! c_g - 5290\, c_A^2\!\cdot\! c_t + 49.8\, c_A^2\!\cdot\! c_V^2 + 7300\, c_A^4 + 196\, c_A^2
\nonumber\\
&&
+ 424\, c_g\!\cdot\! c_t - 6.94\, c_g\!\cdot\! c_V^2 + 1270\, c_g^2 - 18.8\, c_g +3.66\, c_t\!\cdot\! c_V^2
\nonumber\\
&&
   +1120\, c_t^2 - 4.47\, c_t + 43.2\, c_V^4 + 248\, c_V^2 + 476\,,
\nonumber\\
N^{100}_{[1100,1500]}&=&
1050\, c_A^2\!\cdot\! c_g - 2750\, c_A^2\!\cdot\! c_t + 21.1\, c_A^2\!\cdot\! c_V^2 + 4010\, c_A^4 + 65.7\, c_A^2
\nonumber\\
&&
- 90.2\, c_g\!\cdot\! c_t - 2.38\, c_g\!\cdot\! c_V^2 + 1300\, c_g^2 - 4.72\, c_g + 1.03\, c_t\!\cdot\! c_V^2
\nonumber\\
&&
    +529\, c_t^2 + 2.08\, c_t + 16.0\, c_V^4 + 91.1\, c_V^2 + 134\,,
\nonumber\\
N^{100}_{[1500,2500]}&=&
1700\, c_A^2\!\cdot\! c_g - 1630\, c_A^2\!\cdot\! c_t + 8.69\, c_A^2\!\cdot\! c_V^2 + 2430\, c_A^4 + 27.0\, c_A^2
          \nonumber\\
&&
-407\, c_g\!\cdot\! c_t - 1.05\, c_g\!\cdot\! c_V^2 + 2000\, c_g^2 - 0.526\, c_g + 0.134\, c_t\!\cdot\! c_V^2
             \nonumber\\
&&
   + 296\, c_t^2 - 1.76\, c_t + 6.38\, c_V^4 + 36.1\, c_V^2 + 46.3\,,
\nonumber\\
N^{100}_{[2500,5000]}&=&
1170\, c_A^2\!\cdot\! c_g - 382\, c_A^2\!\cdot\! c_t + 1.25\, c_A^2\!\cdot\! c_V^2 + 569\, c_A^4 + 4.82\, c_A^2
          \nonumber\\
&&
-350\, c_g\!\cdot\! c_t - 0.0963\, c_g\!\cdot\! c_V^2 + 2140\, c_g^2 - 0.0120\, c_g - 0.0126\, c_t\!\cdot\! c_V^2
             \nonumber\\
&&
+66.7\, c_t^2 - 0.0583\, c_t + 0.846\, c_V^4 + 4.84\, c_V^2 + 5.37\,.
\label{eq:100fit}
\end{eqnarray}
Similarly to the $14$\,TeV analysis, we have obtained these numbers assuming the identification efficiency for each lepton is $95\%$, summing over all the charge/flavor final states and using the following $K$-factors for each bin~\cite{Azatov:2014jga,Ball:2013bra}
\bea
K=\{1.49, 1.41, 1.41, 1.42, 1.46, 1.49, 1.59\}.
\eea
The $q\bar{q}\to 4\ell$ background yields are, for $1$\,ab$^{-1}$,
\begin{equation}
N_{q\bar{q}}^{100} = \{7.30\cdot 10^4, 2.04\cdot 10^4, 5300, 2410, 918, 447, 92.8\}.
\end{equation}

\section{Boosted Higgs analysis}\label{sec:hj}
For the sake of completeness, in this appendix we give more details on the boosted Higgs projections, which are based on the results of Ref.~\cite{Schlaffer:2014osa}. We concentrate only on the $h\to \tau\tau$ decay. 

\subsection{14\,TeV}
We divide the Higgs transverse momentum distribution in four bins, 
\begin{equation}
p_T = (300, 400, 500, 600, \infty)\;\mathrm{GeV}.
\end{equation}
For each bin, the signal cross section is in general a quadratic polynomial in $c_t,c_g$,
\bea
&\sigma_{14} = \alpha_{14}\, c_t^2 + \beta_{14}\, c_g^2 + \gamma_{14}\, c_t\!\cdot\! c_g,
\eea
where $\alpha_{14}$ is the pure SM cross section, $\beta_{14}$ is the cross section mediated solely by the contact Higgs-gluon interaction, and $\gamma_{14}$ is the interference cross section. The values of $\alpha_{14},\beta_{14}$ and $\gamma_{14}$ were extracted from Tables III and V of Ref.~\cite{Schlaffer:2014osa} and are reported in the second to fourth columns of Table~\ref{tab:hjet_projection}.
\begin{table}
\begin{center}
\begin{tabular}{c|c|c|c|c|c|c}
$p_T \; [\mathrm{GeV}]$ & $\alpha_{14}~[{\rm fb}]$ & $\beta_{14}~[{\rm fb}]$ & $\gamma_{14}~[{\rm fb}]$ & $R_{gg}$ & $R_{qg}$ & $R_{q\bar{q}}$ \\
\hline
$[300,400]$  & $(0.172)_{0.62}$ & $(0.271)_{0.52}$  & $(0.420)_{0.57}$ & 61.5 & 25.3 & 15.4 \\
$[400,500]$  & $(0.052)_{0.58}$ & $(0.117)_{0.47}$  & $(0.150)_{0.53}$ & 83.0 & 30.9 & 17.7 \\
$[500,600]$  & $(0.013)_{0.54}$ & $(0.038)_{0.44}$ & $(0.043)_{0.49}$ & 109 & 37.3 & 20.3 \\
$[600,\infty]$  & $(0.009)_{0.48}$ & $(0.047)_{0.37}$ & $(0.038)_{0.43}$ & 142 & 44.7 & 23.2 \\
\end{tabular}
\end{center}
\caption{\label{tab:partonL} Parameters used to rescale the $14$\,TeV boosted Higgs results of Ref.~\cite{Schlaffer:2014osa} to the FCC. See text for details.}
\label{tab:hjet_projection}
\end{table}
The event yields for $1$\,ab$^{-1}$ are then
\bea
N^{14}_{[300,400]} &=& 172\, c_t^2 +271\, c_g^2 + 420\, c_t\!\cdot\! c_g,\, \nonumber\\
N^{14}_{[400,500]} &=& 52\, c_t^2 + 117\, c_g^2 + 150\, c_t\!\cdot\!  c_g\, ,\nonumber\\
N^{14}_{[500,600]} &=& 13\, c_t^2 + 38\, c_g^2 + 43\, c_t\!\cdot\! c_g\,, \nonumber\\
N^{14}_{[600,\infty]} &=& 9\, c_t^2 + 47\, c_g^2 + 38\, c_t\!\cdot\!  c_g \,,
\eea
whereas the total background is
\bea
N^{14}_{\rm bkg} = (427, 135, 37, 25).
\eea
Following Ref.~\cite{Schlaffer:2014osa}, we assign to the background yield in each bin an uncertainty equal to $(N^{14}_{\rm bkg})^{-1/2}$, which should be thought of as originating from the statistical uncertainty of the background measurement in the sideband regions.

\subsection{100\,TeV}
%
Here we present a simplified method to derive FCC projections from the results of Ref.~\cite{Schlaffer:2014osa}, that consists in rescaling the $14$\,TeV cross sections by the relevant parton luminosity ratios. For this purpose, we need to know the breakdown of the cross sections $\alpha_{14},\beta_{14},\gamma_{14}$ by partonic channel. The fraction of each cross section that comes from the $gg$ initial state~\cite{Azatov:2013xha,Grojean:2013nya} is reported in the second to fourth columns of Table~\ref{tab:hjet_projection} as an underscript, for example $(\alpha_{14})_{x_{gg}^\alpha}$, and similarly for $\beta$ and $\gamma$. Then for a given bin the $100$\,TeV cross section can be estimated as
\begin{align}
\sigma_{100} \,\simeq & \, \alpha_{14} [x_{gg}^{\alpha} R_{gg} + (1-x_{gg}^{\alpha}) R_{qg}] c_t^2 + \beta_{14} [x_{gg}^{\beta} R_{gg} + (1-x_{gg}^{\beta}) R_{qg}] c_g^2 \nonumber \\ 
\,+& \gamma_{14} [x_{gg}^{\gamma} R_{gg} + (1-x_{gg}^{\gamma}) R_{qg}] c_t\!\cdot\! c_g\,,
\end{align}
where we have neglected the small contributions of the $\bar{q}g$ and $q\bar{q}$ partonic channels. The FCC/LHC parton luminosity ratios $R_{gg,\,qg}$ are reported in the fifth and sixth columns of Table~\ref{tab:hjet_projection}. To compute them, for each bin in $p_T$ we have approximated the partonic center of mass energy with the smallest kinematically allowed value, $\hat{s} = m_{h}^{2}+2 \bar{p}_{T}^{2}+2 \bar{p}_{T}\sqrt{\bar{p}_{T}^{2}+m_{h}^{2}}\,$, where $\bar{p}_{T}$ is the lower end of the bin, and taken $\sqrt{\bar{p}_T^2 + m_h^2}$ as factorization scale. The MSTW2008 LO PDFs~\cite{Martin:2009iq} were used. The signal event yields for $1$\,ab$^{-1}$ are then
\bea
\label{eq:fcchj}
N^{100}_{[300,400]} &=& 8230\, c_t^2 + 11900\, c_g^2 + 19400\, c_t\!\cdot\! c_g \,,\nonumber\\
N^{100}_{[400,500]} &=& 3180\, c_t^2 + 6510\, c_g^2 + 8760\, c_t\!\cdot\! c_g \,,\nonumber\\
N^{100}_{[500,600]} &=& 990\, c_t^2 + 2600\, c_g^2 + 3100\, c_t\!\cdot\! c_g\,,\nonumber\\
N^{100}_{[600,\infty]} &=& 820\, c_t^2 + 3800\, c_g^2 + 3300\, c_t\!\cdot\! c_g \,.
\eea
For the background estimation, we have assumed that $WW$+jets and $Z$+jets are produced in $q\bar{q}$ collisions (the parton luminosity ratios for the $q\bar{q}$ channel are given in the last column of Table~\ref{tab:partonL}), whereas $t\bar{t}$+jets is dominated by the $gg$ initial state. This leads to the total background prediction,
\bea
N_{\rm bkg}^{100} = (12000, 4940 , 1200 , 1170).
\eea
The background uncertainty was included in the same way as in the $14$\,TeV analysis.


\end{document}